\documentclass[11pt]{article}

\pdfoutput=1

\usepackage[usenames]{color}
\usepackage[debug,colorlinks=true
,urlcolor=blue
,anchorcolor=blue
,citecolor=green
,filecolor=blue
,linkcolor=blue
,menucolor=blue
,pagecolor=blue
,linktocpage=true,pageanchor=false
]{hyperref}

\usepackage[centertags]{amsmath}
\usepackage{amsfonts}
\usepackage{amssymb}
\usepackage{bbm}
\usepackage{bbold}
\usepackage{amsthm}
\usepackage{newlfont}
\usepackage[a4paper]{geometry}
\usepackage{graphicx}
\usepackage{feynmf}
\usepackage{array}

\numberwithin{equation}{section}

\newcommand{\cA}{\mathcal{A}}
\newcommand{\cN}{\mathcal{N}}
\newcommand{\cF}{\mathcal{F}}

\newcommand{\bR}{\mathbb{R}}
\newcommand{\bZ}{\mathbb{Z}}

\newcommand{\tr}{\mathrm{Tr}}
\newcommand{\Li}{\mathrm{Li}}

\newcommand{\cc}{\mathrm{c.c.}}
\newcommand{\sign}{\mathrm{sign}}

\newcommand{\e}{\epsilon}
\newcommand{\p}{\partial}

\newcommand{\scalar}{\mathrm{scalar}}
\newcommand{\fermion}{\mathrm{fermion}}
\newcommand{\CS}{\mathrm{CS}}

\newcommand{\ec}{\,,}
\newcommand{\ecq}{\ec\quad}
\newcommand{\ed}{\,.}

\newcommand{\YM}{\mathrm{YM}}

\renewcommand{\title}[1]{\vbox{\center\LARGE{#1}}\vspace{5mm}}
\renewcommand{\author}[1]{\vbox{\center#1}\vspace{5mm}}
\newcommand{\address}[1]{\vbox{\center\em#1}}

\def\be{ \begin{equation} }
\def\ee{ \end{equation} }
\def\la#1{\label{#1}}
\def\nref#1{(\ref{#1})}
\def\half{{1\over  2}}
\def\intmesq{ { \cal D}^3 q }

\def\intmesl{{\cal D}^3 l}

\begin{document}

\unitlength = .8mm

\begin{titlepage}
\begin{center}
\hfill \\
\hfill \\

\title{
\vspace{-22mm}
\hfill{\small \tt WIS/18/12-NOV-DPPA}\vskip 50pt
{\ }\\{\ }\\
The Thermal Free Energy in Large $N$ \\ Chern-Simons-Matter Theories}

\renewcommand{\thefootnote}{\fnsymbol{footnote}}
\author{Ofer Aharony$^{a,b}$, Simone Giombi$^c$, Guy Gur-Ari$^b$, \\
 Juan Maldacena$^a$ and Ran Yacoby$^{b}$ }

\address{${}^a$School of Natural Sciences, Institute for Advanced Study, Princeton, NJ 08540, USA}
\address{${}^b$Department of Particle Physics and Astrophysics,
Weizmann Institute of Science,\\ Rehovot 76100, Israel}
\address{${}^c$Jadwin Hall,  Princeton University, Princeton, NJ 08540, USA}
\end{center}

\abstract{
We compute the thermal free energy in large $N$ $U(N)$ Chern-Simons-matter theories with matter fields (scalars and/or fermions) in the fundamental representation, in the large temperature limit. We note that
in these theories the eigenvalue distribution of the holonomy of the gauge field along the thermal circle does not localize even at very high temperatures, and this affects the computation significantly. We verify that our results are consistent with the conjectured dualities between Chern-Simons-matter theories with scalar fields and with fermion fields, as well as with the strong-weak coupling duality of the ${\cal N}=2$ supersymmetric Chern-Simons-matter theory.}

\vfill

\end{titlepage}
\hypersetup{pageanchor=true}

\eject

\tableofcontents

\section{Introduction}

An interesting class of conformal field theories in three dimensions comes from coupling a
(topological) Chern-Simons (CS) theory to massless matter fields. Since the Chern-Simons coupling
is quantized and cannot be renormalized beyond the one-loop level, it is easy to make such
theories conformal by tuning away any relevant operators. When the matter fields are purely
fermionic this is enough to make the theory conformal. When scalar fields are present
there are also classically marginal couplings (of the schematic form $\phi^6$ or
$\phi^2 \psi \psi$) and one has to arrange for their beta functions to vanish, but often
this can be done, at least in the large $N$ limit.

In the past year, a specific class of conformal Chern-Simons-matter theories has been investigated
in detail \cite{Giombi:2011kc,Aharony:2011jz,Chang:2012kt,Jain:2012qi,Aharony:2012nh,fermion_corrs}.
This is the large $N$ 't Hooft limit of a $U(N)$ Chern-Simons theory
at level $k$, coupled to a single matter field (scalar or fermion) in the fundamental
representation (the generalization to an $O(N)$ theory and/or to
$N_f$ matter fields is straightforward). The large
$N$ limit involves taking both $N$ and $k$ to infinity, keeping fixed the ratio $\lambda
\equiv N/k$. These theories come in two versions; a ``regular'' version where one does not
turn on any other couplings except the gauge coupling, and a ``critical'' version which can
be defined by turning on a quartic coupling and sending its coefficient to infinity, or by
adding a mass parameter and making it into a dynamical field that is integrated over.
It was found that this class of theories has some very nice properties :

\begin{itemize}

\item For $\lambda=0$ the theory is free and has an exact high-spin symmetry, with spin
$s$ generators for every positive integer $s$. For finite $\lambda$ and large $N$ this symmetry is still
approximately present, in the sense that the anomalous dimensions of the high-spin currents
are of order $1/N$ \cite{Giombi:2011kc,Aharony:2011jz}.

\item The approximate high-spin symmetry allows for an exact evaluation of all 2-point
and 3-point functions of ``single-trace'' primary operators\footnote{We will refer to
gauge-invariant operators coming from the contraction of a single fundamental and a
single anti-fundamental
field (with any number of covariant derivatives) as single-trace operators; their large
$N$ properties are analogous to those of single-trace operators in theories of adjoint
fields.}, up to two (or three) parameters \cite{Maldacena:2011jn,Maldacena:2012sf}.

\item The 2-point and 3-point correlators of these theories can also be computed exactly in the planar
limit by resumming Feynman diagrams, and agree with the general results based on
symmetries \cite{Aharony:2012nh,fermion_corrs}.

\item It was proposed \cite{Giombi:2011kc, Aharony:2011jz} that these theories (as well as various supersymmetric extensions \cite{Chang:2012kt}) have a gravity dual given 
by a parity breaking version of Vasiliev's higher-spin theory on $AdS_4$ \cite{Vasiliev:1992av, Vasiliev:1999ba}, with a parity-breaking parameter $\theta$ that is a function of the 't Hooft coupling $\lambda$ (the exact gauge theory computation of \cite{Aharony:2012nh}, direct 3-point function calculations in the bulk \cite{Giombi:2012ms} and arguments in \cite{Chang:2012kt} suggest a simple linear relation between the bulk phase $\theta$ and $\lambda$). This duality is a generalization of the original conjectures \cite{Klebanov:2002ja,Sezgin:2003pt} relating the singlet sector of bosonic/fermionic vector models to the parity invariant Vasiliev's theories.


\item The gravity dual is the same for the theory of scalars coupled to Chern-Simons,
and for fermions coupled to Chern-Simons, suggesting a duality between these theories,
and this duality is consistent also with their planar $2$-point and $3$-point functions.
More precisely, the ``critical'' version of the scalar theory has the same gravity dual as the
``regular'' version of the fermionic theory, and vice versa.

\end{itemize}

In the large $N$ limit, the considerations above suggest an exact duality between
the ``critical'' $U(N)_{k_{\YM}}$ CS theory coupled to a scalar field, and the ``regular''
$U(k_{\YM})_{-N}$ CS theory coupled to a fermionic field; here we are using the definition of
the coupling constant $k_{\YM}$ that arises when the theory is regularized using a Yang-Mills
theory at high energies (this is the definition that agrees with the level of the affine
Lie algebra that arises on boundaries of the Chern-Simons theory, but many formulas depend on the
renormalized coupling $k = k_{\YM}+N \,\sign(k_{\YM})$; in this paper we will generally use this renormalized
coupling in all formulas, unless indicated otherwise). In the absence of the matter fields this
duality is simply the level-rank duality of pure Chern-Simons theories \cite{Witten:1988hf, Naculich:1990pa, Camperi:1990dk, Mlawer:1990uv, Naculich:2007nc,Kapustin:2010mh} (and, whenever they are
non-zero, the Chern-Simons contributions dominate in the large $N$ limit over those of the matter fields), but the claim is that there is still an exact duality also after adding the matter fields.
The duality was conjectured in \cite{Aharony:2012nh} to hold also for finite values
of $N$, with a small shift in the CS level on the fermionic side.

The tests of this (non-supersymmetric) duality so far involve exact computations of
various 2-point and 3-point correlation functions in the planar limit, which agree
precisely. However, as shown in \cite{Maldacena:2012sf}, these correlators are all determined by the
almost-conserved high-spin symmetry, and it would be nice to be able to test the duality
by properties that are not determined purely by this symmetry. This is particularly
important since the gravity dual Vasiliev theory appears to contain additional coupling constants
that could affect 5-point and higher correlators of single-trace operators, and their role (if any) in these dualities
is still not clear, so one may wonder whether the two theories may have the same
2-point and 3-point functions for their single-trace operators
(determined by the high-spin symmetry) but still not
be identical.

In this paper we provide further evidence for this duality by matching the thermal
free energy between the two theories mentioned above, at large enough temperatures and
in the planar limit. The
free energy for the fermionic theory was computed already in \cite{Giombi:2011kc},
while the scalar
computation (by the same methods) was performed in \cite{Jain:2012qi, unp}.
These computations do not agree
with the duality (nor with supersymmetric versions of the duality, for which there
is much more evidence); the scalar computation does not even show a vanishing free
energy in the strong coupling limit $\lambda \to 1$ where $\langle T(x) T(y) \rangle$ goes to
zero, indicating that it cannot possibly be correct.

The computations in the literature assumed that the eigenvalue distribution of the
holonomy of the gauge field around the thermal circle localizes around zero at
high enough temperatures, so that the periodicity of the fermionic/scalar fields
can be taken to be the naive one (without any shift coming from the gauge holonomy).
For free theories this is certainly correct \cite{Sundborg:1999ue,Aharony:2003sx};
the matter fields tend to
cause the eigenvalue distribution to localize, and while on some spatial manifolds
(like $S^2$) there is a repulsive term coming from the measure, the effects
of this term go to zero at high temperatures. For asymptotically free gauge
theories it is also clearly true that the holonomy eigenvalues localize at high enough
temperatures. However, for conformal theories at finite coupling it is not clear
if this is correct or not.

In section \ref{holonomy-sec} we argue that the correct large $N$ eigenvalue distribution for the
thermal gauge holonomy in a Chern-Simons-matter
theory of the type described above is not localized, but actually is a uniform
distribution with a width covering a fraction $|\lambda|$ of the unit circle.
Recall that in our convention $\lambda$ goes from zero at weak coupling to one
at strong coupling, so that at weak coupling the eigenvalues do localize as
expected, while in the strong coupling limit the eigenvalue distribution
becomes uniform (as in a confined phase, though there is no confinement here).
This effect comes from the Chern-Simons interactions, that make the
eigenvalues effectively fermionic so that they cannot be too dense, despite the
attractive forces coming from the matter sector.

In sections \ref{prelims}-\ref{fermion_theories} we compute the thermal free energy for the scalar
and fermion theories, both ``regular'' and ``critical'', using this eigenvalue
distribution. The computation is completely analogous to the computations of \cite{Giombi:2011kc,Jain:2012qi,unp},
just with a different periodicity condition for the matter fields, arising from
the holonomy. We show that with this correct eigenvalue distribution the free
energies have the expected properties (going to zero in the strong coupling
limit), and that there is a precise matching between the thermal free energies
of the scalar and fermion theories. This provides an additional test for the
duality between these two theories, that does not follow just from their
approximate high-spin symmetry (as far as we know).

In section \ref{bos-fer}, we then generalize the computation to a theory that has both scalars and fermions, with general couplings. In particular, we verify that in the special case of the theory with $\cN=2$ supersymmetry, the thermal free energy in the large $N$ limit is invariant under the supersymmetric version of the duality \cite{Giveon:2008zn,Benini:2011mf}.

Since this paper is rather long, we summarize our results for the free energies of the various theories in section \ref{summary-sec}. Our analysis of the critical theories automatically gives us the thermal free energies also for the Chern-Simons theories coupled to massive scalars and fermions, and in this section we also briefly discuss some of their properties.

Our methods for computing the holonomy apply to any
Chern-Simons-matter theory in which the number of matter fields is much
smaller than $N^2$ in the large $N$ limit.  It would be interesting to understand
their implications for theories like those of \cite{Aharony:2008ug} where the number of matter fields
is of order $N^2$, so that one cannot neglect their back-reaction on the
eigenvalue distribution. This should be important in any computation of the
(planar) thermal free energy in these theories.
Our methods are sufficient for
computing the order $\lambda^2$ correction to the free energy in these theories.

It would be interesting to understand how to reproduce our results for the thermal
free energy (or various
other effects of the Chern-Simons part of the conformal field theory \cite{Banerjee:2012gh,Banerjee:2012aj})
in the gravity dual theory.
It would also be interesting to find additional tests for the field theory duality, in particular
for finite values of $N$.

\section{The Holonomy Along the Thermal Circle}
\label{holonomy-sec}

When one studies the thermodynamics of gauge theories it is common to
have long distance  modes that require special treatment
 \cite{LindeTS,Svetitsky:1985ye,Braaten:1994qx}.
These are captured cleanly by thinking of the thermal computation as a
  Kaluza-Klein reduction of the Euclidean theory on a circle. If we go to energies  lower than
the inverse radius of the circle (the temperature), we might encounter light degrees
of freedom that contribute to the free energy. In particular,  the holonomy of the gauge
field on the circle is classically a massless field. These light degrees of freedom
need to be taken into account correctly.

In our case, we are interested in conformal field theories coming from
large $N$ $U(N)$ Chern-Simons theories coupled to matter fields in the fundamental
representation. On general grounds, the free energy of a 3d conformal field theory at
temperature $T$ and volume $V$ is proportional, for large enough volumes, to $V T^2$.
For a standard gauge theory coupled to fields in the fundamental representation, the
large $N$ free energy would be dominated by the contribution of the gauge fields
(scaling as $N^2$), while the matter field contributions scale as $N$. In our case,
the Chern-Simons theory is purely topological, so the contributions of this theory
on its own are independent of the volume, and one thus expects the free energy to
scale as $N$. One may expect that at large enough volumes (or, equivalently, large
enough temperatures) the contribution of the matter fields will be dominant, and
will make the eigenvalues of the holonomy matrix of the gauge field localize near
zero (as in asymptotically free gauge theories). This is the assumption that was
made in previous computations of the free energy. However, we will see that the
presence of the Chern-Simons theory modifies this, even at very large volumes.

In order to analyze the thermal partition function,
we reduce the Euclidean theory on the thermal circle of circumference $\beta = 1/T$.
For the purposes of this section,
we begin by integrating out the matter fields in the fundamental representation (at finite temperature
and coupling we expect all these fields to be massive).
We are then left with an effective two dimensional theory, involving the
dimensional reduction of the gauge field. We have the two dimensional gauge fields, and also the
holonomy of the gauge field along the thermal circle,
 \be
a \equiv i \int_0^\beta  dx^3 A_3.
\ee
The $i$ is introduced so that $a$ is hermitian (we use a convention in which the $A_\mu$
fields are anti-Hermitian).

Let us first consider this dimensional reduction for the pure Euclidean Chern-Simons theory,
\begin{align}
  S_{\CS} &= \frac{ik}{4\pi} \int \! d^3x \, \e^{\mu\nu\rho} \,
  \tr\left( A_\mu \p_\nu A_\rho - \frac{2i}{3} A_\mu A_\nu A_\rho\right)
  \ecq k \in \bZ \ed
  \label{SCSzero}
\end{align}
In that case we get an effective
action of the form
\be \la{twodred}
S = { k  \over 2 \pi } \int \! d^2 x \, \tr(F a),
\ee
where $a $ is the holonomy (which is a scalar field in two dimensions),
and $F$ is the field strength along the two non-compact dimensions.
Since the original action is topological, this action is also topological.

Let us now compactify one of the $\mathbb{R}^2$ directions on a circle of circumference $L$,
and think of the other non-compact direction as (Euclidean) time. We now have a second holonomy,
\be
b \equiv i \int_0^L dx^1  A_1.
\ee
We can gauge fix the other component of the gauge field $A_0$ to zero, and
the two dimensional action then becomes a quantum mechanics action for the
two holonomies, of the form
\be
S = { i k  \over 2\pi }\int \! dt \, \tr(a \partial_0 b).
\ee

This has the form of a matrix quantum mechanics action with a single matrix degree of freedom, since $b$ is canonically conjugate to $a$. We can diagonalize $a$ and consider only the dynamics of
 the eigenvalues. As is standard in matrix quantum mechanics, this introduces a Vandermonde determinant from the measure, which we can swallow into the wavefunction, but this makes the remaining part of the wavefunction anti-symmetric in the eigenvalues, so they need to be treated as identical (due to the
 remaining Weyl symmetry) fermions \cite{Brezin:1977sv}. Thus, we get a Euclidean action for $N$ fermionic variables \cite{Douglas:1994ex},
\be
\label{fermact}
S = { i k  \over 2 \pi }  \sum_{i=1}^N  \int \! dt \, \dot b_i a_i.
\ee
For each eigenvalue we have   the action for a particle in a magnetic field, on the lowest Landau level. Due to the
fact that $a_i$ and $b_i$ come from holonomies, they all have periodic identifications, which in
 our normalization are $a_i \sim a_i + 2 \pi$ and $b_i \sim b_i + 2 \pi $. So we have $N$ fermions moving on a torus, with a magnetic field that has $k$ units of flux. Having flux $k$ implies that each
Landau level has precisely $|k|$ states.
So, we have $N$ fermions filling $|k|$ states. In our conventions $k$ is the renormalized Chern-Simons coupling (which for the $SU(N)$ part of the gauge group is related to the Chern-Simons coupling that arises from a Yang-Mills regularization by $k = k_{\YM} + N {\mathrm{sign}}(k_{\YM})$), so we always have $|k| >  N$, and we have
 ${|k| \choose N}$ different ways of filling these states\footnote{Note that there are other regularizations in which one gets a similar action (\ref{fermact}), but with a coefficient $k_{\YM}$ (for the $SU(N)$ part) and with bosonic eigenvalues \cite{Elitzur:1989nr}.
Note that there are no  Vandermonde determinant factors
 for the Chern-Simons theory on $T^3$ \cite{Blau:1993tv},
 but the eigenvalues still behave as fermions in this case \cite{Douglas:1994ex}.}.
 In the pure Chern-Simons theory the Hamiltonian vanishes, so all these states have the same energy (zero energy), giving the ground state degeneracy
 of the $U(N)$ Chern-Simons theory on a spatial torus \cite{Douglas:1994ex}.

Up to now our analysis was exact. Next, we add the matter in the fundamental representation of the gauge group, in the large $N$ limit. We imagine that the thermal circle is
much smaller than the other circle, and we expect all the matter fields to obtain masses of the order of
the temperature. When we integrate out the matter fields we get a contribution to the free energy of order $N$ times the
volume. In this computation (which we will perform in detail in the subsequent
sections for a specific holonomy) we treat the holonomy as a constant background field, so we obtain a result which
depends on the holonomy. In the limit where the thermal circle is much smaller than the other
circles, the other effects of the gauge field are very small. The most important term that we get
 in the effective action is then a potential for the holonomy
\be \la{potgen}
\int dt dx^1 V(a ),
\ee
where $V(a)$ is some (gauge-invariant) functional of the holonomy, which scales as $N$ in the
large $N$ limit. (Additional terms depending on the two dimensional
gauge fields also arise when integrating out
the matter fields, and we will discuss them below.)
At least to leading order in the $1/k$ expansion, the potential $V(a)$ has a minimum at $a=0$, and the expansion around zero looks like $V(a)-V(0) \propto   \tr(a^2) T^2  $.

\begin{figure}[!ht]
\centering
  \includegraphics[width=0.6\textwidth]{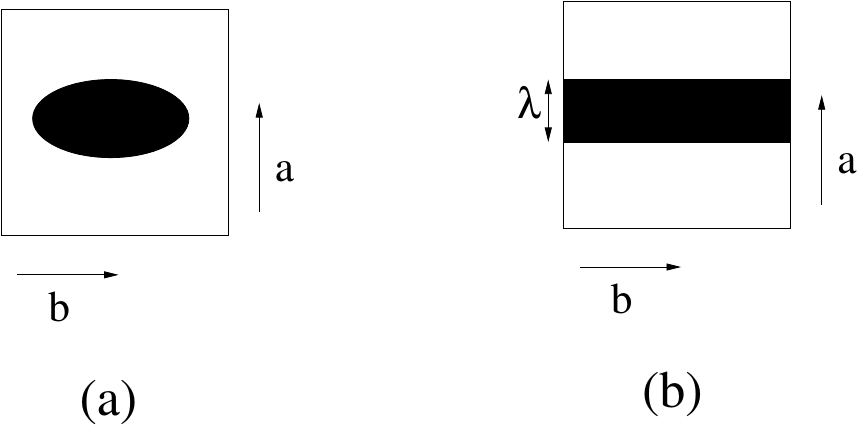}
  \caption{({\bf a}) One out of many possible  configurations of fermions in the lowest Landau level on the torus.
  ({\bf b}) The potential generated by integrating out the matter fields selects the configuration where the
 fermions are localized in a band around $a=0$. The fact that $L \gg \beta$ breaks the symmetry between the
 two holonomies. }
  \label{droplets}
\end{figure}

One may be tempted to set $a=0$, as was done in previous computations in thermal Chern-Simons-matter theories. However, recall that
 when we consider this theory compactified on an extra $S^1$, we saw that the eigenvalues of the
holonomy $a$ behave like the positions of fermions in the lowest Landau level. Thus, we cannot set them all to zero;
the state $a_i =0$ is not consistent with the Pauli exclusion principle. The best we can do is to fill a Fermi sea
so that the $N$ levels are as close to zero (in the $a$ direction) as possible. Since $k a_i / 2\pi$ is the conjugate momentum to $b_i$, which has periodicity $2\pi$, if we work in a basis of eigenstates of $a_i$, then the eigenvalues are
quantized in units of $(2\pi / |k|)$. We want to fill the $N$ eigenstates that are closest to $a=0$.
In the ``phase space'' torus parameterized by $a$ and $b$, this means that we are filling
all states in a band of width $ 2 \pi |\lambda|$ around $a=0$, and uniform in the $b$ direction,
see figure \ref{droplets}(b).
Notice that in the large $N$ limit, we have many eigenvalues and we can view the
 distribution of fermions as a fluid, ignoring
the details of the wavefunctions of these fermionic eigenvalues on the torus.

The matter contributions break the degeneracy between the different ground states of the Chern-Simons theory, and imply that this specific state has the lowest energy. We will argue below that the other
contributions from integrating out the matter fields do not affect this computation at leading order
in the large $N$ limit. In principle we should now sum over all the different states with a thermal
distribution. The energy density difference between the lowest energy state that we found and the next state is of order $T^3 / |k|$ (since we can shift a single eigenvalue by $2\pi/k$). Thus, if
$VT^2$ is much larger than $N$ and $k$, we can focus on the ground state, and ignore all other
contributions to the free energy.
From here on we will be working in this limit for simplicity.
The regime $VT^2 \sim 1$ was discussed in \cite{Banerjee:2012gh,Banerjee:2012aj}.

The discussion above implies that at large volumes/temperatures
we should diagonalize the thermal holonomy and  distribute the eigenvalues uniformly
in a band of width $ 2 \pi |\lambda|$ around
zero. It is important that even though to derive this result we compactified an extra circle,
this width is independent of the radius of the second spatial circle that we used for this derivation, as
 long as $L \gg \beta$.
 So, this distribution of holonomy eigenvalues affects the extensive part of the free energy (the part
that is proportional to the radius of the second circle).
 It turns out that this change in the eigenvalue distribution affects the thermal free energy already at order $\lambda^2$ in the weak coupling expansion. As a consistency check on this value for the holonomy, one can check that with this eigenvalue distribution the traces of the holonomy matrix $e^a$ transform in the correct way under the level-rank duality (the traces of this holonomy in symmetric representations in the theory with coupling $\lambda$ are mapped to the traces of the holonomy in anti-symmetric representations in the theory with coupling
 $\lambda-{\mathrm{sign}}(\lambda)$).
 This would not be the case for $a=0$.

It seems surprising that a finite-volume effect like the eigenvalues being fermions modifies the
free energy even in the large volume limit. We can obtain
an alternative perspective on this by following the standard treatment
 of  long distance modes in thermal gauge theories \cite{Braaten:1994qx}, adapted for Chern-Simons matter
 theories. Again, we consider for simplicity the large $k$ limit (with large $N$, and with $\lambda = N/k$ fixed and small).
Then, after dimensionally reducing on the circle we get the following effective action
\be
S =  \int d^2x  \left[ {k \over 2\pi} \tr(F a) + T^2 \tr(a^2)  + \cdots \right],
\ee
where we expanded around $a =0$ and kept just the leading terms, ignoring coefficients of order one.
We can now integrate out the scalar field $a$, obtaining
\be
S \propto {  k^2 \over T^2 } \int \! d^2x \, \tr(F^2).
\ee
This is a two dimensional Yang-Mills action with an effective coupling $g^2_{2D} \propto { T^2 \over k^2 } $.
So for large $k$ this
is very weakly coupled at the scale $T$. In two dimensional Yang-Mills there are no propagating degrees of freedom, but there are non-trivial
flux excitations. Here we are interested in the ground state energy of this effective two dimensional Yang-Mills theory.
 We again compactify    the $x^1$ direction on a circle of circumference $L$. Choosing  the other  direction to be time,
we end up with a matrix quantum mechanics for the holonomy $b = i \int_0^L dx^1 A_1$ around the $x^1$ direction. Again we have free fermions by diagonalizing the matrix; this
time we have free fermions on a circle. The effective Lagrangian takes the
form
\be
S \propto \int dt { 1 \over L } {    k^2 \over T^2 } \sum_{i=1}^N \dot b_i^2.
\ee
Since $b_i$ is periodic with period $2\pi$, the quantized momenta for $b_i$ have the  form
$p_i = n_i$ with integer $n_i$, where $p_i \sim { k^2 \over L T^2 } \dot b_i$. We fill
the Fermi sea up to $|n| = N/2$ and we end up with a ground state energy of the form
\be
E_0 \propto \sum_{n_i=-N/2}^{N/2} {L T^2 \over k^2} n_i^2 \propto { L T^2 N^3 \over   k^2} \propto { L T^2  N \lambda^2  }.
\ee
In particular, we see that it is proportional to $L$, so it is an extensive contribution. And, we see that it is (at leading order in perturbation theory) an order
$\lambda^2$ correction to the leading order in $N$ result for the free energy (that would come
from the vacuum energy of the matter fields themselves); thus, it cannot be ignored in the large $N$ limit.
As usual, in order
to speak in a meaningful way about the ground state energy we need to define a subtraction scheme. The fact that the
scheme we used here is the correct one is most transparent according to our previous discussion, where we started from the
dimensional reduction of Chern-Simons theory and viewed the same correction to the energy as arising from integrating out the   matter
fields in the fundamental representation.

As we integrate out matter fields, we also induce terms of the form
\be \la{otherkin}
\int \! d^2x \,\tr\!\left((D a )^2\right) +  \beta^2 \tr\!\left(F^2\right),
\ee
with order one coefficients (depending on $\lambda$) and with powers of the temperature fixed by dimensional analysis. Such terms do not affect
the previous discussions, because they are subleading in the $1/N$ expansion.

The spreading of the eigenvalues of the holonomy is expected to be a general feature in the
thermodynamics of Chern-Simons-matter theories. For a finite number of matter fields in the
fundamental representation we saw that the eigenvalues have a uniform distribution with a
width proportional to $|\lambda|$, with corrections that are small at large $N$. It is more
complicated to analyze what happens when the number of matter fields is of order $N^2$;
in that case corrections like \nref{otherkin} are no longer negligible, and could affect the
eigenvalue distribution. The precise eigenvalue distribution could also depend on the
precise field content, since (for instance) for adjoint matter fields the potential $V(a)$ depends only on
differences between eigenvalues. In any case, we expect that there should be a non-trivial
eigenvalue distibution also in such cases, which would give extra
terms starting at order $\lambda^2$ in the thermal partition function. In particular, these
effects should be added to
the results in \cite{Smedback:2010ji}, and to any other perturbative computations of the
thermodynamics of Chern-Simons-matter  theories.

The derivation we have presented here is most clear if we consider the free energy for the theory on
$T^2 \times S_\beta^1 $. In this case the $T^2$ gives us the extra circle that we needed to run our argument, and we can take the large volume limit of the $T^2$ and obtain the extensive part of the free energy as we discussed above.
One could wonder if similar effects also appear when we consider the three dimensional theory on $\Sigma_2 \times S_\beta^1$ for other Riemann surfaces $\Sigma_2$,
in the limit that $S^1_\beta$ is very small. Pure three dimensional Chern-Simons theory on such manifolds was studied in detail in \cite{Blau:1993tv}. They argued that the theory can be reduced to a two dimensional theory of
the form \nref{twodred}, where now the integral is over $\Sigma_2$. In addition,  we can further diagonalize the
holonomy in the circle direction. The resulting effective action is independent of the manifold, up to overall factors (involving powers of the Vandermonde determinant) which do not scale like the
volume of $\Sigma_2$, and thus do not contribute to the free energy in the limit of large $\Sigma_2$ volume.
Since the analysis on the torus gave us an extensive effect, we expect that the same effect should arise on a general Riemann surface $\Sigma_2$. It would be nice to show in detail how this works. A particularly interesting
case is when $\Sigma_2 = S^2$. It was argued in \cite{Shenker:2011zf} (for $\lambda =0$)
 that the thermal circle needs to be
very small, of order $1/\sqrt{N}$ as compared to the size of the $S^2$, in order to obtain an answer for the free energy that is linear in $N$. Note that here we are also discussing very large temperatures of the same order. On the sphere, there is a volume independent contribution to the effective action
for the thermal holonomy, $a$,  that leads to a repulsion between eigenvalues. This will lead
to an eigenvalue distribution that is uniformly spread over the whole circle for small temperatures, but for very large temperatures
this is overwhelmed by the effects we encountered on the torus. For negatively curved $\Sigma_2$
the extra volume independent factors lead to attractive interactions between the eigenvalues, so we
expect the eigenvalues to localize already at smaller temperatures.

In the remainder of this paper we will compute the resulting free energy exactly as a function of $\lambda$, but to leading order in the
$1/N$ expansion.
We work on $\bR^2$, which can be thought of as the large volume limit of $T^2$ (as
we mentioned, we expect the result to be the same for large enough volumes on
any two dimensional space).
The computation is similar to those done in \cite{Giombi:2011kc,Jain:2012qi,unp}.
We simply need to repeat all their steps, but now with a constant holonomy $a$ matrix, with the
eigenvalue distribution that we found, in which the
  eigenvalues of the holonomy are distributed in a band as in figure \ref{droplets} (b).
We view this as a constant background field in the perturbative computation of the free energy.

\section{Preliminaries}
\label{prelims}

In the following sections we consider the theory of $N$ complex scalars or Dirac fermions, in the fundamental representation of $U(N)$, coupled to a $U(N)$ gauge field $A_\mu$ with Chern-Simons interactions in 3 Euclidean dimensions. The Chern-Simons action \eqref{SCSzero} can be written as\footnote{The generators $T^a$ are anti-Hermitian, with the normalization $\tr(T^a T^b) = -\frac{1}{2} \delta^{ab}$.}
\begin{align}
  S_{\CS} &= - \frac{ik}{8\pi} \int \! d^3x \, \e^{\mu\nu\rho}
  \left( A_\mu^a \p_\nu A_\rho^a + \frac{1}{3} f^{abc} A_\mu^a A_\nu^b A_\rho^c
  \right)
  \ecq k \in \bZ
  \ed
  \label{SCS}
\end{align}
We take the 't Hooft large $N$ limit, keeping $\lambda = \frac{N}{k}$ fixed,
and work (following \cite{Giombi:2011kc}) in light-cone gauge in the $x^1-x^2$ directions, $A_- = 0$.\footnote{Light-cone coordinates are defined by $x^{\pm} = (x^1 \pm i x^2) / \sqrt{2}$. The invariant tensors are given by $\delta_{33} = \delta_{+-} = 1$ and $\e_{+-3} = i$.} In this gauge the gluon self-interaction vanishes, and the Chern-Simons action reduces to
\begin{align}
  S_{CS} = \frac{k}{4\pi} \int \! d^3x \, A_+^a \p_- A_3^a \ed
  \label{Scs-lc}
\end{align}
We will regularize our theory by placing a sharp momentum cutoff $\Lambda$ in the $x^1-x^2$ directions. This will cure most UV divergences, and the remaining ones (that will occur when computing determinants) will be handled as they are encountered. In this regularization (used also in \cite{Giombi:2011kc,Aharony:2011jz,Jain:2012qi,Aharony:2012nh}), $k$ is related to the level $k_{\YM}$ that appears in a regularization by a Yang-Mills action (and that appears, for instance, in the Wess-Zumino-Witten model on boundaries) by $k = k_{\YM} + N {\mathrm{sign}}(k_{\YM})$ (for the $SU(N)$ part).

We will take our manifold to be $\bR^2 \times S^1$, where the compactified direction is $x^3$ and has length $\beta$. As argued in Section \ref{holonomy-sec}, we should take the gauge field to have non-trivial holonomy around $S^1$ (the eigenvalues of the holonomy are gauge-invariant). We can describe the holonomy in terms of the zero mode
\begin{align}
  \cA_3 = \frac{1}{V_2} \int_{\bR^2} \! d^2x \, A_3 \ec
\label{zero-mode}
\end{align}
where $V_2$ is the regularized volume of $\bR^2$. Using a residual gauge transformation (that preserves the light-cone gauge) we set $\p_3 \cA_3 = 0$ and diagonalize $\cA_3$. 
In the notations of Section \ref{holonomy-sec}, the holonomy along the thermal circle is
\begin{align}
  a \equiv i \beta \cA_3 \ec
\end{align}
whose eigenvalues are real and live on a circle of length $2\pi$, due to the periodic identifications which follow from the gauge symmetry.

Most of the calculation can be done for general fixed $a$, but in the end we are interested in the large $N$ limit where the eigenvalues of $a$ are uniformly spread around $0$ with width $2\pi|\lambda|$, namely we take
\begin{align}
   a_{ii} \to a(u) = 2 \pi |\lambda| u \ecq
  u \in \left[ -\frac{1}{2}, \frac{1}{2} \right] \,,
  \label{alpha}
\end{align}
and replace sums over eigenvalues by integrals over $u$ with uniform distribution over the spread of eigenvalues. In other words, we replace
\begin{align}
\sum_{i=1}^N f(a_{ii}) \longrightarrow {  N } \int_{- \half}^\half d u f( 2 \pi |\lambda| u ) \ed
\end{align}

\section{Bosonic Theories}

In this section we consider (following \cite{Aharony:2011jz,Aharony:2012nh}) the theory of $N$ complex scalars $\phi$, coupled to the gauge field $A_\mu$, defined by the action
\begin{align}
  S &= S_{\scalar} + S_{\CS} \ec \label{action-sc} \\
  S_{\scalar} &= \int \! d^3x \,\, (D_\mu \phi)^\dagger (D^\mu \phi)
  + \frac{\lambda_6}{3! N^2} (\phi^\dagger \phi)^3 \ecq
  D_\mu = \p_\mu + A_\mu \ed
  \label{Sscalar}
\end{align}
We include the coupling $\lambda_6$ which is exactly marginal in the large $N$ limit \cite{Aharony:2011jz}.
We treat the holonomy as part of the free theory. Its effect is to shift the scalar momentum. Let us define
\begin{align}
  \tilde{p}_\mu = p_\mu - i \cA_\mu \ec
\end{align}
where $\cA_3$ is the zero-mode (\ref{zero-mode}) and $\cA_\pm \equiv 0$.
 We will diagonalize the matrix $\cA_3$ so that we have $\tilde p_3 = p_3 - a_{ii}/\beta$ for the
 $i$'th scalar.
 The Feynman rules for this theory in momentum space include (in the $A_-=0$ gauge)
\begin{center}
  \begin{tabular}{m{3cm}l}
    \includegraphics[width=0.2\textwidth]{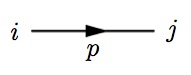}
    & $\displaystyle = ( \tilde{p}^{-2} )_{ji} $
    \\
    \includegraphics[width=0.2\textwidth]{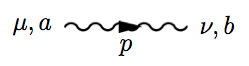}
    & $\displaystyle = G_{\nu\mu}(p) \delta_{ab}$
    \\
    \includegraphics[width=0.2\textwidth]{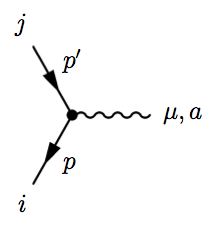}
    & $\displaystyle = i ( T^a \tilde{p}'_\mu + \tilde{p}_\mu T^a )_{ij}$
    \\
    \includegraphics[width=0.2\textwidth]{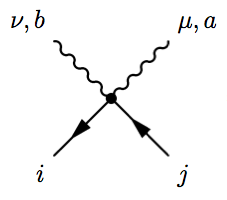}
    & $\displaystyle = \{ T^a, T^b \}_{ij} \delta_{\mu 3} \delta_{\nu 3}$
  \end{tabular}
\end{center}
The non-vanishing components of the gluon propagator are
\begin{align}
  G_{+3}(p) = -G_{3+}(p) = \frac{4\pi i}{k} \frac{1}{p^+} \ed
\end{align}
It is important that this gluon propagator does not depend on $p_3$. Thus, it is not affected by
the holonomy.

\subsection{Exact propagator}
\label{exact-prop-sec}

The exact scalar propagator is\footnote{For the thermal theory $p_3 = { 2 \pi n / \beta}$, $q = 2 \pi n' / \beta$, and
we should interpret
$ 2 \pi \delta( p_3 - q_3) \to \beta \delta_{n,n'}$. }
\begin{align}
  \langle \phi(p)^j \phi^\dagger(-q)_i \rangle =
  \left[ \frac{1}{\tilde{p}^2 - \Sigma_B} \right]_{\!ji}
  (2\pi)^3 \delta^3(p-q)
  \ec
  \label{exact-prop-sc}
\end{align}
where $(\Sigma_B)_{ji}$ is the sum of 1PI diagrams. As we will see, at large $N$ all contributions to $(\Sigma_B)_{ji}$ are proportional to the unit matrix in color space, so $\Sigma_B(p)_{ji} = \Sigma_B(p) \delta_{ji}$. Also, none of the contributions will depend on $p_3$. Using this and the fact that $\Sigma_B$ must be invariant under rotations in the $x^1-x^2$ plane, we see that $\Sigma_B$ can depend on $p$ only through $p_s^2 = (p_1)^2 + (p_2)^2 = 2 p^+ p^-$. Using dimensional analysis, we can write
\begin{align}
  \Sigma_B(p) = - \beta^{-2} \mu_B^2(\lambda, \lambda_6, \beta p_s, a) \ec
  \label{Sigma-and-g-sc}
\end{align}
and $\mu_B \beta^{-1}=\mu_B T$ is the thermal mass of the scalar.\footnote{From now on, we may also refer to the dimensionless quantity $\mu_B$ as the ``thermal mass".}

\begin{figure}[!ht]
\centering
  \includegraphics[width=1.1\textwidth]{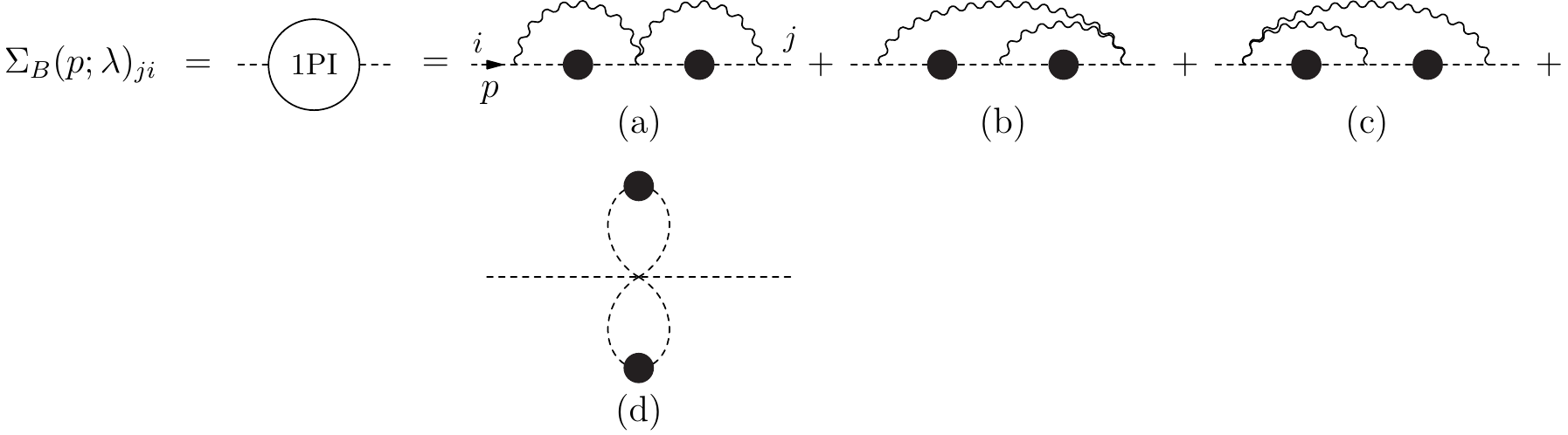}
  \caption{Bootstrap equation for the scalar self-energy. A filled circle denotes the full scalar propagator.
  (d) corresponds to the insertion of the $\lambda_6 (\phi^2)^3$ interaction vertex.}
  \label{fig:bs-scalar}
\end{figure}

The bootstrap equation for $\Sigma_B$ is shown in figure \ref{fig:bs-scalar}. The 1-loop diagram (not shown since it vanishes) can be written as
\begin{align}
  - \int \intmesq
  \left[
  (T^a \tilde{q}^\nu + \tilde{p}^\nu T^a)
  \frac{1}{\tilde{q}^2 - \Sigma_B(q)}
  (T^a \tilde{p}^\mu + \tilde{q}^\mu T^a)
  \right]_{ji} G_{\mu\nu}(q-p) \ed
  \label{1loop-sc}
\end{align}
It is understood here and below that the full integration measure on $\bR^2 \times S^1$ is
\begin{align}
  \int \intmesq =
  \frac{1}{\beta} \int_{\bR^2} \frac{d^2q}{(2\pi)^2} \sum_{n=-\infty}^{\infty} \ec
\end{align}
where momenta $q$ on $S^1$ are quantized as $q_n = \frac{2\pi n}{\beta}$. It is easy to check that the diagram \eqref{1loop-sc} vanishes using the relation for generators in the fundamental representation of $U(N)$,
\begin{align}
  T^a_{ij} T^a_{kl} = -\frac{1}{2} \delta_{il} \delta_{jk} \ec
  \label{fund-rel}
\end{align}
and the fact that $\Sigma_B$ and $\tilde{p}$ are diagonal in color space for any vector $p$.

In the remaining diagrams, the $\phi \partial_\mu \phi^\dagger A_\mu$ vertex always appears with $\mu\ne 3$, and therefore the Feynman rule for the vertex becomes simply $i T^a_{ij} (p+p')_\mu$. The remaining diagrams can be written as
\begin{align}
  \mathrm{(a)} &=
  - \left( \frac{4\pi}{k} \right)^2
  \int
  \intmesl
  \intmesq
  \frac{(l+p)^+}{(l-p)^+}
  \frac{(q+p)^+}{(q-p)^+}
  \left[
  T^b \frac{1}{\tilde{l}^2 - \Sigma_B(l)}
  \{ T^a, T^b \}
  \frac{1}{\tilde{q}^2 - \Sigma_B(q)}
  T^a
  \right]_{ji}
  \ec
  \notag \\
  \mathrm{(b)} = \mathrm{(c)} &=
  \left( \frac{4\pi}{k} \right)^2
  \int
  \intmesl
  \intmesq
  \frac{(l+p)^+}{(l-p)^+}
  \frac{(q+l)^+}{(q-l)^+}
  \left[
  T^a
  \frac{1}{\tilde{l}^2 - \Sigma_B(l)}
  T^b
  \frac{1}{\tilde{q}^2 - \Sigma_B(q)}
  \{ T^a, T^b \}
  \right]_{ji}
  \ec
  \notag \\
  \mathrm{(d)} &=
  - \frac{1}{2} \frac{\lambda_6}{N^2}
  \int
 \intmesl
 \intmesq
  \left[ \frac{1}{\tilde{q}^2 - \Sigma_B(q)} \right]_{j_1 i_1}
  \left[ \frac{1}{\tilde{l}^2 - \Sigma_B(l)} \right]_{j_2 i_2}
  \times
  \notag \\ &\quad \qquad \qquad \;\;
  \left[ \delta_{ij} \delta_{i_1 j_1} \delta_{i_2 j_2} + (5~\mathrm{permutations}) \right]
  \ed
\end{align}
In diagram (d) the one-half is a symmetry factor.
For general matrices $A,B$ we have, using \eqref{fund-rel},
\begin{align}
\label{traceeq}
  T^b A \{ T^a, T^b \} B T^a =
  T^b A T^a B \{ T^a, T^b \} =
  \frac{1}{4} [BA + \tr(A) \tr(B) \mathbb{1} ] \ed
\end{align}
Using this in diagrams (a)-(c), we see that the first term of (\ref{traceeq}) is subleading at large $N$ and can be dropped. In diagram (d), only the first permutation survives at large $N$. The diagrams can therefore be written as
\begin{align}
  \mathrm{(a)} &=
  - \frac{4\pi^2}{k^2} \delta_{ji}
  \left[
  \int
 \intmesq
  \frac{(q+p)^+}{(q-p)^+}
  \tr \left(
  \frac{1}{\tilde{q}^2 - \Sigma_B(q)}
  \right)
  \right]^2
  \ec
  \label{a-diag-sc} \\
  \mathrm{(b)} = \mathrm{(c)} &=
  \frac{4\pi^2}{k^2} \delta_{ji}
 \int
 \intmesl \intmesq
  \frac{(l+p)^+}{(l-p)^+}
  \frac{(q+l)^+}{(q-l)^+}
  \tr \left(
  \frac{1}{\tilde{q}^2 - \Sigma_B(q)}
  \right)
  \tr \left(
  \frac{1}{\tilde{l}^2 - \Sigma_B(l)}
  \right)
  \ec
  \notag \\
  \mathrm{(d)} &=
  - \frac{\lambda_6}{2} \delta_{ji} \left[ \frac{1}{N}
  \int
  \intmesq
  \tr \left( \frac{1}{\tilde{q}^2 - \Sigma_B(q)} \right)
  \right]^2
  \ed \notag
\end{align}
We can simplify things by taking the derivative of the bootstrap equation
\begin{align}
  \Sigma_B(p) \delta_{ij} = \mathrm{(a)} + \mathrm{(b)} + \mathrm{(c)}
  + \mathrm{(d)}
\end{align}
with respect to $p^-$, using the relations
\begin{align}
  \frac{\p}{\p p^-} \frac{1}{p^+} = 2\pi \delta^2(p) \ecq
  \frac{\p p_s}{\p p^-} = \frac{p^+}{p_s} \ed
  \label{p-diff}
\end{align}
The resulting equation is $\p_{p^-} \Sigma_B(p) = 0$, and therefore $\Sigma_B$ and $\mu_B^2$ do not depend on $p_s$. Using this observation, we can simply evaluate the various contributions to $\Sigma_B$ by setting $p_s=0$. With this choice, one finds that (b)=(c)=0 for symmetry reasons, and
\begin{align}
\mathrm{(a)} &=
  - 4\pi^2\lambda^2 \delta_{ji}
  \left[
  \int^{1/2}_{-1/2} du \int
  \intmesq
  \frac{1}{\tilde{q}^2 +\beta^{-2}\mu_B^2}
  \right]^2\ec \\
\mathrm{(d)} &=
  -\frac{\lambda_6}{2} \delta_{ji}
  \left[
  \int^{1/2}_{-1/2} du \int
  \intmesq
  \frac{1}{\tilde{q}^2 +\beta^{-2}\mu_B^2}
  \right]^2\,,
\end{align}
where we have replaced the trace with an integral over the uniform spread of eigenvalues. The sum over the $q_3$ momentum takes the form
\begin{align}
  F(q_s,u) &\equiv \frac{1}{\beta} \sum_{n=-\infty}^\infty
  \frac{1}{(\frac{2\pi}{\beta}(n+|\lambda|u))^2 + q_s^2+\beta^{-2}\mu_B^2} \notag\\
  &= \frac{\beta}{2\sqrt{(\beta q_s)^2 + \mu_B^2}}
 \cdot {\mathrm{Re}} \left( \coth
  \left[ \frac{\sqrt{(\beta q_s)^2 + \mu_B^2}+2\pi i |\lambda|u}{2} \right] \right) \, \ed
\end{align}
The bootstrap equation for the self-energy can then be written as
\begin{equation}
  \Sigma_B =
  - \hat{\lambda}^2
  \left[\int^{1/2}_{-1/2}du \int dq_s q_s F(q_s,u) \right]^2
  \label{bs-sc}
  \ec
  \end{equation}
  where we defined
  \begin{equation}
  \hat{\lambda}^2 \equiv \lambda^2 + \frac{\lambda_6}{8\pi^2}
  \ed
\end{equation}
The thermal mass $\mu_B$ then obeys the bootstrap equation
\begin{align}
  |\mu_B(\lambda,\lambda_6)| =
  \left| \frac{\hat{\lambda}}{2}
  \int^{1/2}_{-1/2} du
  \int_0^{\infty} dx \frac{x}{\sqrt{x^2+\mu_B^2}}
  \, {\mathrm{Re}} \left( \coth \left[
  \frac{\sqrt{x^2 + \mu_B^2} +2\pi i |\lambda|u}{2}
  \right] \right) \right| \ed
  \label{bs-g-sc}
\end{align}
In writing this we made a change of variables $x = \beta q_s$, and took the square root of the equation for $\Sigma_B$.
Using a cutoff $\Lambda'=\beta\Lambda$, we find that the radial integral gives (using ${\mathrm{Re}} (\log (z)) = \log |z|$)
\begin{align}
  2 \log \left| \sinh \left( \frac{\sqrt{x^2+\mu_B^2} +2\pi i |\lambda|u}{2} \right)
  \right|_{x=0}^{\Lambda'}
  =
  \Lambda'
  - 2 \log \left| 2 \sinh \left( \frac{|\mu_B| +2\pi i |\lambda|u}{2} \right) \right|
  + O \left( \frac{1}{\Lambda'} \right)
  \ed
\end{align}
Subtracting the divergence using a mass counter-term, and taking the limit $\Lambda \to \infty$, the bootstrap equation for $\mu_B$ now reads
\begin{align}
  | \mu_B | &=
  \left| \hat{\lambda}
  \int^{1/2}_{-1/2} du
  \log \left| 2 \sinh
  \left( \frac{|\mu_B| +2\pi i |\lambda|u}{2} \right) \right| \right| \ed
\label{muB-u}
\end{align}

This integral can be computed analytically in terms of dilogarithms, and we find\footnote{This is easily done by again replacing $\log |z| = {\mathrm{Re}} (\log (z))$.} the equation\footnote{In this paper the dilogarithm function is defined on the principal branch, with a cut from $1$ to $+\infty$ along the real axis.}
\begin{align}
  \pm \mu_B
  &=
  - \frac{|\hat{\lambda} \mu_B|}{2} - \frac{1}{2 \pi i}
  \frac{|\hat{\lambda}|}{|\lambda|}
  \left[ \Li_2(e^{-|\mu_B|-\pi i |\lambda|}) - \Li_2(e^{-|\mu_B|+\pi i |\lambda|}) \right]
  \label{mass-nc-sc}
  \ed
\end{align}
In some cases there may be more than one solution to this equation (with different choices of
the relative sign), and in such cases we should choose
the solution that minimizes the free energy. We can always take (without loss
of generality) $\mu_B > 0$, and we will do this in the rest of this section.
The solution that minimizes the free energy is then always the
one with the plus sign in \eqref{mass-nc-sc}.

To summarize, the exact scalar propagator is given by
\begin{align}
  \langle \phi(p)^j \phi^\dagger(-q)_i \rangle =
  \left[ \frac{1}{\tilde{p}^2 + \beta^{-2} \mu_B^2(\lambda,\lambda_6)} \right]_{\!ji}
  (2\pi)^3 \delta^3(p-q)
  \ec
\end{align}
where the thermal mass $\mu_B$ is found by solving \eqref{mass-nc-sc}. Note that if we drop the integral over the holonomy in (\ref{muB-u}) and set $u=0$, the equation determining the thermal mass agrees with the result derived in \cite{Jain:2012qi,unp}.

\subsection{Free energy}

It was shown in \cite{Jain:2012qi,unp} that the free energy has a simple expression in terms of the self-energy. In our case the calculation follows through, except that we should include the integral over the holonomy.  
At large $N$, we have
\begin{align}
  \beta F_B = N V_2 \int_{-1/2}^{1/2} du
  \int \frac{d^2p}{(2\pi)^2} \sum_{n=-\infty}^\infty
  \left[
  \log \left( \tilde{p}^2 - \Sigma_B \right)
  + \frac{2}{3} \frac{\Sigma_B}{\tilde{p}^2 - \Sigma_B}
  \right] \ec
  \label{F-sc-1}
\end{align}
where  $p_3 = \frac{2\pi n}{\beta}$ and $\tilde p_\mu = p_\mu -  { 2 \pi |\lambda| u \over \beta } \delta_{3,\mu}$.

Consider first the log term. The sum over $n$ diverges, and we regulate it by introducing a negative contribution with a large mass $\beta^{-1} M$, so that we need to compute (here we use that $\Sigma_B=-\beta^{-2}\mu_B^2$)
\begin{align}
  \int \frac{d^2p}{(2\pi)^2} & \sum_{n=-\infty}^\infty
  \log \left[
  \frac{ \tilde{p}^2 +\beta^{-2}\mu_B^2 }{ \tilde{p}^2 + \beta^{-2} M^2 }
  \right]
  =
  \int \frac{d^2p}{(2\pi)^2}
  \log \prod_{n=-\infty}^\infty \left[
  1 + \frac{p_s^2+\beta^{-2}\mu_B^2}{\left( \frac{2\pi}{\beta}(n-|\lambda|u) \right)^2}
  \right]
  - (\mu_B \to M)
  \notag \\ &=
  \int \frac{d^2p}{(2\pi)^2}
  \log \left[
  \cosh(\beta \sqrt{p_s^2 +\beta^{-2}\mu_B^2}) - \cos(2\pi |\lambda|u)
  \right]
  - (\mu_B \to M)
  \notag \\ &=
  \frac{1}{2\pi \beta^2} \int_0^{\Lambda'} \! dx \, x
  \log \left[
  2 \cosh(\sqrt{x^2 + \mu_B^2}) - 2 \cos(2\pi |\lambda|u)
  \right]
  - (\mu_B \to M)
  \ec
  \label{F-logterm-sc}
\end{align}
where again $x=\beta p_s$ and $\Lambda' = \beta \Lambda$. The first term has a UV power divergence, which we now subtract. We can write this term as
\begin{align}
  \frac{1}{2\pi \beta^2} \int_0^{\Lambda'} \! dx \, x \left\{
  \sqrt{x^2+\mu_B^2} + \log \left[
  1 + e^{-2\sqrt{x^2+\mu_B^2}} - 2 e^{-\sqrt{x^2+\mu_B^2}} \cos(2\pi |\lambda|u)
  \right]
  \right\} \ed
\end{align}
The divergence is now contained in the first term. The last term, which is finite, can be simplified by taking $y = \sqrt{x^2 + \mu_B^2}$ and sending $\Lambda \to \infty$. After subtracting the divergence we find
\begin{align}
  \frac{1}{2\pi \beta^2}
  \left[
  - \frac{\mu_B^3}{3}
  + \int_{\mu_B}^\infty \! dy \, y
  \log \left(
  1 + e^{-2y} - 2 e^{-y} \cos(2\pi |\lambda|u)
  \right)
  \right] \ed
\end{align}
Let us now compute the integral $\int_{-1/2}^{1/2} du$ of this expression. With some work, one finds the result\footnote{
The following relation is useful here:
\begin{align}
  \Li_2\left( e^{2\pi i x} \right) +
  \Li_2\left( e^{-2\pi i x} \right) = 2\pi^2 B_2(x) \ecq
  0 < {\mathrm{Re}} \, (x) < 1 \ec
\end{align}
where $B_2(x) = x^2 - x + \frac{1}{6}$ is the second Bernoulli polynomial.
}
\begin{align}
  \frac{1}{2\pi \beta^2}
  \left\{
  - \frac{\mu_B^3}{3}
  - \frac{1}{\pi i |\lambda|}
  \int_{\mu_B}^\infty \! dy \, y
  \left[
  \Li_2 (e^{-y+\pi i |\lambda|}) - \Li_2 (e^{-y-\pi i |\lambda|})
  \right]
  \right\}
  \ed
  \label{F-term1-sc}
\end{align}
This is the first term in \eqref{F-logterm-sc}. The second, regulator term in \eqref{F-logterm-sc} gives a pure UV divergence (when taking $M,\Lambda\to\infty$) that we subtract. This removes the determinant regulator.

Let us now consider the second term in \eqref{F-sc-1}. Here the computation is similar to what we encountered in section \ref{exact-prop-sec}. The bootstrap equation for the self-energy implies that
\begin{align}
\mu_B = \pm 2\pi |\hat\lambda| \int_{-1/2}^{1/2} du \int \frac{d^2p}{(2\pi)^2} \sum_{n=-\infty}^\infty
  \frac{1}{\tilde{p}^2 +\beta^{-2}\mu_B^2} \ec
\end{align}
and so we find that
\begin{align}
  \frac{2}{3} \int_{-1/2}^{1/2} du \int \frac{d^2p}{(2\pi)^2} \sum_{n=-\infty}^\infty
  \frac{\Sigma_B}{\tilde{p}^2 - \Sigma_B}
  =\mp \frac{\mu_B^3}{3\pi\beta^2|\hat\lambda|}
  \ed
  \label{F-term2-sc}
\end{align}

Combining \eqref{F-term1-sc} and \eqref{F-term2-sc}, we find that the free energy is given by
\begin{align}
  \beta F_B = - \frac{N V_2}{2\pi\beta^2} \left\{
  \frac{\mu_B^3}{3}
  \left( 1 \pm \frac{2}{|\hat{\lambda}|} \right)
  + \frac{1}{\pi i |\lambda|}
  \int_{\mu_B}^\infty \! dy \, y
  \left[
  \Li_2 (e^{-y+\pi i |\lambda|}) - \Li_2 (e^{-y-\pi i |\lambda|})
  \right]
  \right\}
  \ec
  \label{F-nc-sc}
\end{align}
where the choice of sign is correlated to the choice of sign in \eqref{mass-nc-sc}.
Note that \eqref{mass-nc-sc} is always given by setting to zero the $\mu_B$-derivative of
\nref{F-nc-sc}.
Using \eqref{mass-nc-sc} we can rewrite \nref{F-nc-sc}  as
\begin{align}
  \beta F_B = - \frac{N}{\lambda} \frac{V_2}{2\pi^2 i \beta^2}
  &
  \left\{
  \frac{\mu_B^2}{3}
  \left[
  \Li_2 (e^{-\mu_B+\pi i \lambda}) - \Li_2 (e^{-\mu_B-\pi i \lambda})
  \right]
  \right.
  \notag \\
  &
  \quad \left.
  + \int_{\mu_B}^\infty \! dy \, y
  \left[
  \Li_2 (e^{-y+\pi i \lambda}) - \Li_2 (e^{-y-\pi i \lambda})
  \right]
  \right\}
  \ed
  \label{F-nc-sc-1}
\end{align}
It is obvious from this expression that the free energy vanishes (for fixed $N$) in the
strong coupling limit $|\lambda| \to 1$.
The free energy with $\lambda_6=0$ is shown in figure \ref{fig:F-sc} below.

\subsection{The critical theory}

In this subsection we consider the critical bosonic theory, which at large $N$ can be defined as the Legendre transform of the regular theory (with $\lambda_6=0$) with respect to the scalar operator $\phi^\dagger \phi$. In order to carry out the transform, we begin by computing the free energy in the theory deformed by a mass squared $\sigma$, and then we make $\sigma$ into a dynamical field. The mass-deformed theory has the action \eqref{action-sc}, deformed by
\begin{align}
  \delta S = \int \! d^3x \, \sigma \phi^\dagger \phi \ed
\end{align}
The exact propagator \eqref{exact-prop-sc} is now
\begin{align}
  \langle \phi(p)^j \phi^\dagger(-q)_i \rangle =
  \frac{1}{\tilde{p}^2 + \sigma - \Sigma_B}
  \delta_{ji} (2\pi)^3 \delta^3(p-q)
  =
  \frac{1}{\tilde{p}^2 - \Sigma_{B,\sigma}}
  \delta_{ji} (2\pi)^3 \delta^3(p-q)
  \ed
\end{align}
The computation of the bootstrap equation for the self-energy in section \ref{exact-prop-sec} generalizes to this case with minor modifications. The right-hand side of equation \eqref{bs-sc}, given by the sum of 1PI diagrams, is given by the old one with the replacement $\Sigma_B \to \Sigma_{B,\sigma}$. If we replace the definition \eqref{Sigma-and-g-sc} with $\Sigma_{B,\sigma} = - \beta^{-2} \mu_B^2$, the right-hand side is unchanged. In the left-hand side of \eqref{bs-sc} we now have
$\Sigma_B = \Sigma_{B,\sigma} + \sigma = - \beta^{-2} (\mu_B^2 - \tilde{\sigma})$, where $\tilde{\sigma} \equiv \beta^2 \sigma$.
As a result, the final bootstrap equation \eqref{mass-nc-sc} becomes
\begin{align}
  \pm \sqrt{\mu_B^2 - \tilde{\sigma}} =
  - \frac{|\lambda| \mu_B}{2} - \frac{1}{2 \pi i}
  \left[ \Li_2(e^{-\mu_B-\pi i |\lambda|}) - \Li_2(e^{-\mu_B+\pi i |\lambda|}) \right] \ed
  \label{mass-crit-sc-1}
\end{align}

The free energy written in terms of the self-energy is also slightly modified, and becomes
\begin{align}
  \beta F_{B,\sigma} = N V_2 \int_{-1/2}^{1/2} du \int \frac{d^2p}{(2\pi)^2} \sum_{n=-\infty}^\infty
  \left[
  \log \left( \tilde{p}^2 - \Sigma_{B,\sigma} \right)
  + \frac{2}{3} \frac{\Sigma_{B,\sigma} + \sigma}{\tilde{p}^2 - \Sigma_{B,\sigma}}
  \right] \ed
\end{align}
The extra contribution in the last term can be easily computed analogously to \eqref{F-term2-sc}. The final result for the free energy is given by
\begin{align}
  \beta F_{B,\sigma} = - \frac{N V_2 T^2}{2\pi} \left\{
  \frac{\mu_B^3}{3}
  \pm \frac{2 (\mu_B^2 - \tilde{\sigma})^{3/2}}{3 |\lambda|}
  + \frac{1}{\pi i |\lambda|}
  \int_{\mu_B}^\infty \! dy \, y
  \left[
  \Li_2 (e^{-y+\pi i |\lambda|}) - \Li_2 (e^{-y-\pi i |\lambda|})
  \right]
  \right\}
  \ed
  \label{bs-F-sc}
\end{align}

\begin{figure}
  \centering
  \includegraphics[width=0.6\textwidth]{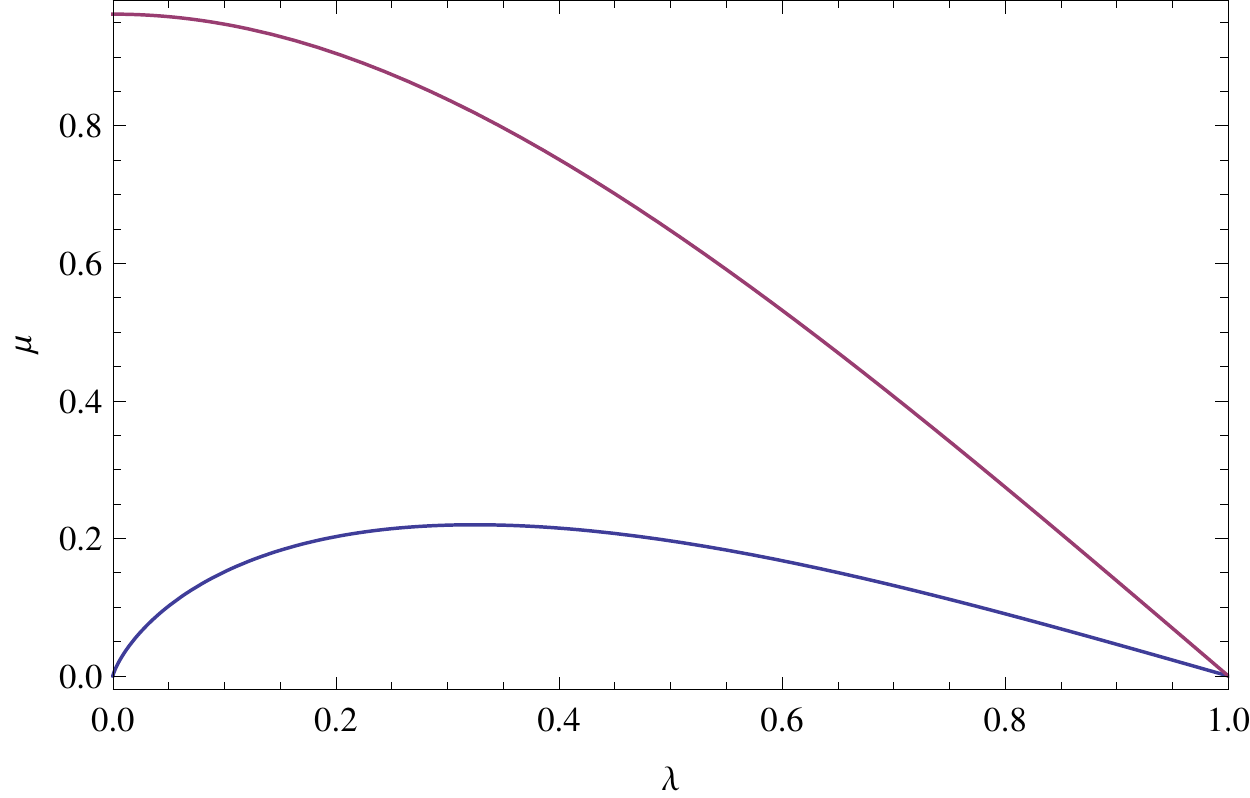}
  \caption{Thermal mass in the regular bosonic theory with $\lambda_6=0$
  (blue, lower curve) and critical theory (red, upper curve), with arbitrary normalization.}
  \label{fig:g-sc}
\end{figure}
\begin{figure}
  \centering
  \includegraphics[width=0.6\textwidth]{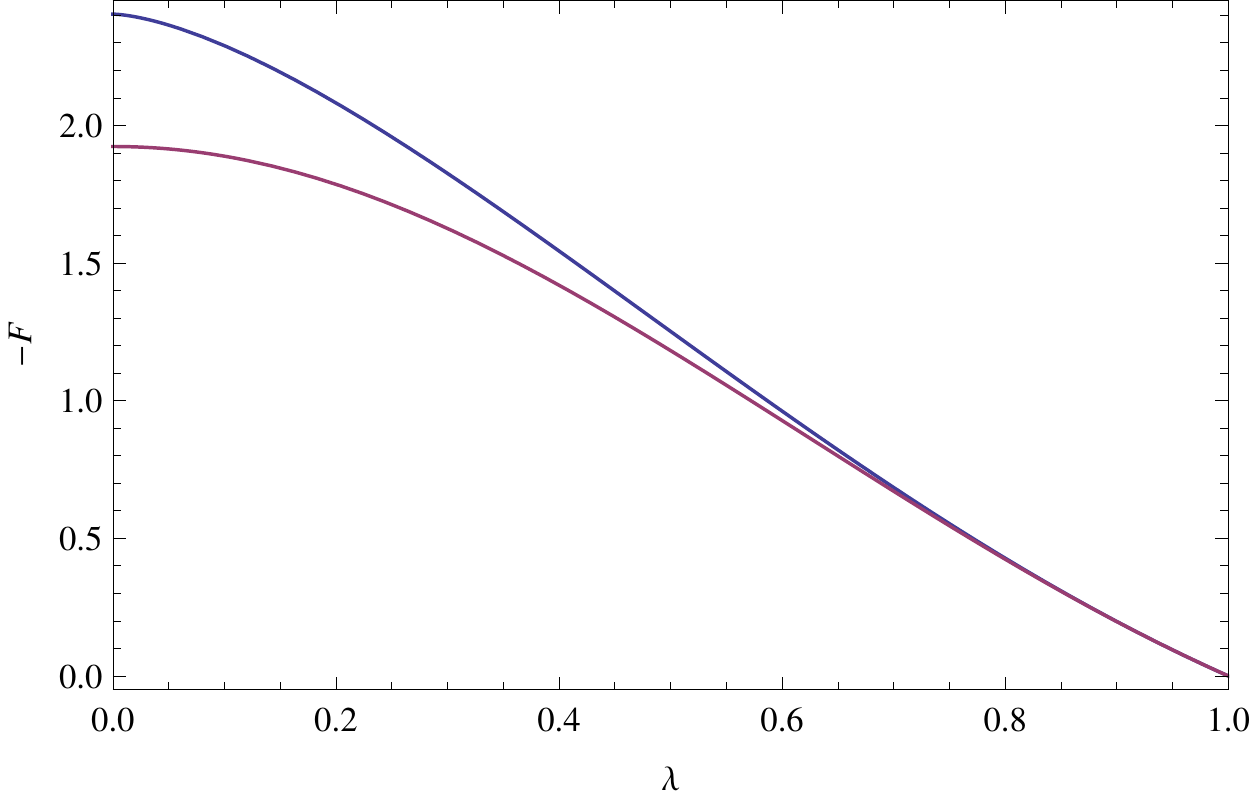}
  \caption{Negative free energy in the regular bosonic theory with
  $\lambda_6=0$ (blue, upper curve) and critical theory (red, lower curve), using arbitrary
  normalization.}
  \label{fig:F-sc}
\end{figure}

Let us now carry out the path integral over $\sigma$ to obtain the critical theory.
The saddle point equation for $\tilde{\sigma}$, derived from \eqref{bs-F-sc}
(using \eqref{mass-crit-sc-1}),
is
\begin{align}
  \sqrt{\mu_B^2 - \tilde{\sigma}} = 0 \ec
\end{align}
and we find $\tilde{\sigma} = \mu_B^2$. Plugging this back in, we find the free
energy of the critical bosonic theory,
\begin{align}
  \beta F_B^{\mathrm{crit.}} =
  - \frac{N V_2 T^2}{2\pi} \left\{
  \frac{\mu_{B,c}^3}{3}
  + \frac{1}{\pi i \lambda}
  \int_{\mu_{B,c}}^\infty \! dy \, y
  \left[
  \Li_2 (e^{-y+\pi i \lambda}) - \Li_2 (e^{-y-\pi i \lambda})
  \right]
  \right\}
  \ec
  \label{F-crit-sc}
\end{align}
where $\mu_{B,c}$ is given by (\ref{mass-crit-sc-1})
\begin{align}
  0 =
  \lambda \mu_{B,c} + \frac{1}{\pi i}
  \left[ \Li_2(e^{-\mu_{B,c}-\pi i \lambda}) - \Li_2(e^{-\mu_{B,c}+\pi i \lambda}) \right] \ed
  \label{mass-crit-sc}
\end{align}
These results are shown in figures \ref{fig:g-sc} and \ref{fig:F-sc}.
It is amusing to note that  \nref{mass-crit-sc} is simply given by setting the
$\mu_{B,c}$ derivative of \nref{F-crit-sc} to zero.

One may check that in the limit $\lambda=0$ these results agree with the thermal mass and free energy of the critical $O(N)$ model computed in \cite{Sachdev:1993pr}. At $\lambda=0$, one finds from (\ref{mass-crit-sc})-(\ref{F-crit-sc})
\begin{align}
&\mu_{B,c}=2\log \left(\frac{\sqrt{5}+1}{2} \right)\label{On-crit-mass} \ec \\
&\beta F_{B,c}=-N V_2 T^2\,\frac{4}{5}\frac{\zeta(3)}{\pi}\,,
\end{align}
which is indeed the result of \cite{Sachdev:1993pr} (up to a factor of 2 in the free energy, due to the fact that we work with $U(N)$ rather than $O(N)$).

\section{Fermionic Theories}
\label{fermion_theories}

In this section we consider, following \cite{Giombi:2011kc}, the theory of $N$ Dirac fermions $\psi^i$, coupled to the $U(N)$ Chern-Simons gauge field $A_\mu$. The theory is defined by the action
\begin{align}
  S &= S_{\fermion} + S_\CS , \\
  S_{\fermion} &=
  \int \! d^3 x \,
  \bar{\psi} \gamma^\mu D_\mu \psi + \sigma \bar{\psi} \psi \ed
  \label{Sfermion}
\end{align}
The regular theory is given by setting $\sigma=0$. In the critical theory (the Gross-Neveu model with Chern-Simons interactions) $\sigma$ plays the role of the primary scalar operator, and we must perform the path integral over it.

The calculation is similar to the one done in \cite{Giombi:2011kc}. Our conventions for the gamma matrices are $\gamma^\mu = \sigma^\mu$, $\mu=1,2,3$, where $\sigma^\mu $
 are the Pauli matrices. In the thermal theory the fermions have anti-periodic boundary conditions on the thermal $S^1$ (before including the effects of the holonomy), so the momenta in this direction are quantized as $p_3 = \frac{2\pi}{\beta} (n + \frac{1}{2})$, $n \in \bZ$.

\subsection{Exact propagator}

\begin{figure}
  \centering
  \includegraphics[width=0.6\textwidth]{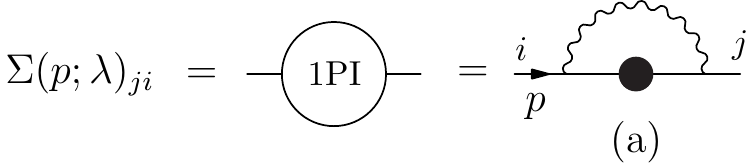}
  \caption{Bootstrap equation for $\Sigma_F$. A filled circle denotes the exact
  fermion propagator.}
  \label{fig:bs-fermion}
\end{figure}

Denote the exact propagator as
\begin{align}
  \langle \psi(p)_i \bar{\psi}(-q)_j \rangle =
  \left[
  \frac{1}{i \tilde{p}_\mu \gamma^\mu + \sigma - \Sigma_F(p)}
  \right]_{ji} (2\pi)^3 \delta^3(p-q) \ed
  \label{exact-prop-fer}
\end{align}
The notation is the same as for the bosonic theories, with $\Sigma_F$ being the sum of 1PI diagrams. The bootstrap equation for $\Sigma_F$ is shown in figure \ref{fig:bs-fermion}.
The $2 \times 2$ matrix $\Sigma_F$ can be uniquely expanded as the linear combination
\begin{align}
  \Sigma_F &= i \Sigma_\mu \gamma^\mu + \Sigma_I I
  = i \Sigma_- \gamma^- +
  i \Sigma_+ \gamma^+ + i \Sigma_3 \gamma^3 + \Sigma_I I
  \ed
\end{align}
Let us also define $\tilde{\Sigma}_I = \Sigma_I - \sigma$.
The diagram in the bootstrap equation is given by\footnote{Let $\gamma^{[\alpha|} A \gamma^{|\beta]} \equiv \gamma^\alpha A \gamma^\beta - \gamma^\beta A \gamma^\alpha$.}
\begin{align}
  \mathrm{(a)} &= \frac{4\pi i}{k}
  T^a_{jm} T^a_{ki}
  \int \!
  \intmesq
  \,
  \frac{1}{(q-p)^+}
  \, \gamma^{[3|}
  \left( \frac{1}{i \tilde{q}_\mu \gamma^\mu + \sigma - \Sigma_F(q)} \right)_{mk}
  \gamma^{|+]}
  \notag \\ &=
  - \frac{2\pi i}{k} \delta_{ji} \tr_N
  \int \!
  \intmesq
   \,
  \frac{1}{(q-p)^+}
  \, \gamma^{[3|}
  \frac{1}{i \tilde{q}_\mu \gamma^\mu + \sigma - \Sigma_F(q)}
  \gamma^{|+]}
  \notag \\ &=
  \frac{2\pi i}{k} \delta_{ji} \tr_N
  \int \!
  \intmesq
  \,
  \frac{1}{(q-p)^+}
  \frac{1}{(\tilde{q} - \Sigma_F)^2 + \tilde{\Sigma}_I^2}
  \, \gamma^{[3|} \!
  \left[
  i (\tilde{q} - \Sigma_F)_\mu \gamma^\mu + \tilde{\Sigma}_I I
  \right]
  \! \gamma^{|+]}
  \ec
\end{align}
where in the second line we used \eqref{fund-rel}. The right-hand side is proportional to the unit matrix in color space, so we may write $\Sigma_F(p)_{ji} = \Sigma_F(p) \delta_{ji}$.
Using the relations $\gamma^{[3|} \gamma^\mu \gamma^{|+]} = -2 \delta^\mu_- I$ and $[\gamma^3,\gamma^+] = 2\gamma^+$, the bootstrap equation becomes
\begin{align}
  \Sigma_F(p) &=
  \frac{4\pi i}{k} \tr_N
  \int \!
  \intmesq
  \,
  \frac{1}{(q-p)^+}
  \frac{\tilde{\Sigma}_I(q) \gamma^+ - i (q-\Sigma_F)^+ I}
  {(\tilde{q} - \Sigma_F)^2 + \tilde{\Sigma}_I^2}
  \notag \\ &=
  4\pi i \lambda
  \int_{-1/2}^{1/2} \! du \,
  \int \! \frac{d^2q}{(2\pi)^2} \,
  \frac{1}{(q-p)^+}
  \frac{1}{\beta}
  \sum_{n=-\infty}^{\infty}
  \frac{\tilde{\Sigma}_I(q) \gamma^+ - i (q-\Sigma_F)^+ I}
  {(\tilde{q} - \Sigma_F)^2 + \tilde{\Sigma}_I^2}
  \label{bs-inter-fer}
  \ed
\end{align}
We see that $\Sigma_- = \Sigma_3 = 0$, and that $\Sigma_F$ is independent of $p^3$.
Using also rotation invariance in the $p^1-p^2$ plane, we may therefore write
\begin{align}
  \tilde{\Sigma}_I(p) = f(\lambda, \beta p_s, \tilde{\sigma}) p_s \ecq
  \Sigma_+(p) = g(\lambda, \beta p_s, \tilde{\sigma}) p_+ \ec
\end{align}
where $\tilde{\sigma} \equiv \beta \sigma$.
Let us now compute the sum in \eqref{bs-inter-fer}. As before, let $x = \beta q_s$.
\begin{align}
  \frac{1}{\beta}
  \sum_{n=-\infty}^{\infty}
  & \frac{1}{(\tilde{q} - \Sigma_F)^2 + \tilde{\Sigma}_I^2(q)}
  =
  \frac{1}{\beta}
  \sum_n
  \frac{1}{\tilde{q}_3^2 + (q-\Sigma_F)_s^2 + \tilde{\Sigma}_I^2}
  \notag \\ &=
  \beta \sum_n
  \frac{1}{ \left[ 2\pi\left( n + \frac{1}{2} - |\lambda|u\right)\right]^2
  + x^2 (1 - g + f^2) }
  \notag \\ &=
  \frac{\beta}{2}
  \frac{1}{x \sqrt{1 - g(x) + f^2(x)}}
  \, {\mathrm{Re}} \left[ \tanh \left( \frac{x \sqrt{1 - g(x) + f^2(x)} +2\pi i |\lambda| u}{2} \right) \right]
  \notag \\ &\equiv
  \frac{\beta}{2} G(x)
  \ed
  \label{G-fr}
\end{align}
The bootstrap equation is now (comparing separately the coefficients of $\gamma^+$ and $I$)
\begin{align}
  \Sigma_+(p) &=
  2\pi \lambda \beta
  \int_{-1/2}^{1/2} du \,
  \int \! \frac{d^2q}{(2\pi)^2} \,
  \frac{\tilde{\Sigma}_I(q)}{(q-p)^+}
  \cdot G(x)
  \ec \label{bs-SigPlus-fer} \\
  \Sigma_I(p) &=
  2\pi \lambda \beta
  \int_{-1/2}^{1/2} du \,
  \int \! \frac{d^2q}{(2\pi)^2} \,
  \frac{q^+}{(q-p)^+}
  \cdot G(x)
  \label{bs-SigI-fer}
  \ed
\end{align}
Let us differentiate with respect to $p^-$, using \eqref{p-diff}. We find
\begin{align}
  \frac{y}{2} \p_y g + g(y) &= - y f G(y) \ec \\
  \p_y f + \frac{f(y)}{y} &= -G(y) \ed
\end{align}
Combining these, we find that
\begin{equation}
  \p_y (y^2 g - y^2 f^2) = 0
  \quad \Longrightarrow \quad
  g(\lambda,y) - f^2(\lambda,y) = - \frac{\mu_F^2(\lambda)}{y^2} \ec
  \label{gfc}
\end{equation}
where $\mu_F$ is an integration constant that may depend on the coupling.
As we will see, $\mu_F$ plays the role of the fermion's thermal mass.
Returning to \eqref{bs-SigI-fer}, we can write it as
\begin{align}
  y f(y) + \beta \sigma &=
  \frac{\lambda}{2\pi}
  \int_{-1/2}^{1/2} du
  \int_0^{\Lambda'} dx \, x G(x)
  \int_0^{2\pi} d\theta_q \,
  \frac{q^+}{(q-p)^+}
  \notag \\ &=
  \lambda
  \int_{-1/2}^{1/2} du
  \int_y^{\Lambda'} dx \,
  \frac{x}{\sqrt{x^2 + \mu_F^2}}
  \, {\mathrm{Re}} \left[ \tanh \left( \frac{\sqrt{x^2 + \mu_F^2} +2\pi i |\lambda|u}{2} \right) \right]
  \notag \\ &\to
  \lambda \Lambda'
  - 2\lambda
  \int_{-1/2}^{1/2} du \,
  \, {\mathrm{Re}} \left( \log \left[
  2 \cosh \left( \frac{\sqrt{y^2+\mu_F^2} +2\pi i |\lambda|u}{2} \right)
  \right] \right)
  \ed
\end{align}
Here again $\Lambda' = \beta \Lambda$ ($\Lambda$ being the cutoff).
In the second line we computed the angular integral, which evaluates to $2\pi \Theta(q_s - p_s)$, and substituted $G(x)$. In the third line we took $\Lambda \to \infty$, keeping diverging terms. We subtract the linear divergence with a mass counter-term. For a trivial holonomy ($u=0$) we find agreement with \cite{Giombi:2011kc}. For our holonomy the remaining integral can be computed, and we find
\begin{align}
  y f(y) + \tilde{\sigma} &=
  - \lambda \sqrt{y^2 + \mu_F^2}
  + \frac{1}{\pi i}
  \left[
  \Li_2 \left( - e^{-\sqrt{y^2 + \mu_F^2} + \pi i \lambda} \right)
  - \cc
  \right] \ed
  \label{f-fer}
\end{align}
To determine the integration constant $\mu_F$ in \eqref{gfc} (which we choose to satisfy $\mu_F > 0$), we demand that $g$ is regular in the limit $y \to 0$. The expression $f^2 - \frac{\mu_F^2}{y^2}$ has a $y^{-2}$ divergence in this limit, unless $\mu_F$ satisfies the equation
\begin{align}
  \pm \mu_F(\lambda) =
  \tilde{\sigma} + \lambda \mu_F + \frac{1}{\pi i}
  \left[
  \Li_2 \left( - e^{-\mu_F-\pi i \lambda} \right) - \cc
  \right] \ed
  \label{mass-nc-fer}
\end{align}
It turns out that this equation always has a unique solution (with some choice of sign).

Let us summarize. The exact fermion propagator in the thermal theory is
\begin{align}
  \langle \psi(p)_i \bar{\psi}(-q)_j \rangle &=
  \frac{1}{i \tilde{p}_\mu \gamma^\mu - i g p_+ \gamma^+ - f p_s I}
  \delta_{ji} (2\pi)^3 \delta^3(p-q)
  \notag \\ &=
  \frac{- i \tilde{p}_\mu \gamma^\mu + i g p_+ \gamma^+ - f p_s I}
  {\tilde{p}^2 + T^2 \mu_F^2}
  \delta_{ji} (2\pi)^3 \delta^3(p-q) \ec
\end{align}
where $g$ is given by \eqref{gfc}, $f$ by \eqref{f-fer}, and $\mu_F$ obeys \eqref{mass-nc-fer}. Notice that $T |\mu_F|$ is the thermal mass.

\subsection{Free energy}
\label{sec:F-fer}

For trivial holonomy, it was shown in \cite{Giombi:2011kc} that the free energy $F$ has a simple expression in terms of the self-energy. In our case the expression is similar, with the same modifications that were needed in the scalar case \eqref{F-sc-1}:
\begin{align}
  \beta F_F &= - N V_2 \int_{-1/2}^{1/2} du
  \int \! \frac{d^2q}{(2\pi)^2} \sum_{n=-\infty}^{\infty}
  \tr_\mathrm{fer.}
  \left\{
  \log\left( i \tilde{q}_\mu \gamma^\mu - \tilde{\Sigma}_F(q) \right)
  + \frac{1}{2} \Sigma_F(q)
  \frac{1}{ i\tilde{q}_\mu \gamma^\mu - \tilde{\Sigma}_F(q) }
  \right\}
  \ec
  \label{F-fr-1}
\end{align}
where $\tilde{\Sigma}_F \equiv i \Sigma_\mu \gamma^\mu + \tilde{\Sigma}_I I
= \Sigma_F - \sigma I$.
Let us first compute the trace over the fermionic indices :
\begin{align}
  \tr_\mathrm{fer.}
  & \left\{
  \log\left( i \tilde{q}_\mu \gamma^\mu - \tilde{\Sigma}_F(q) \right)
  + \frac{1}{2} \Sigma_F(q)
  \frac{1}{ i\tilde{q}_\mu \gamma^\mu - \tilde{\Sigma}_F(q) }
  \right\}
  \notag \\ &=
  \log \left[ \det\,\!\!_{\mathrm{fer.}} \!\! \left( i \tilde{q}_\mu \gamma^\mu - \tilde{\Sigma}_F(q) \right) \right]
  + \frac{1}{2}
  \tr_\mathrm{fer.} \!\! \left(
  \Sigma_F(q)
  \frac{i (\tilde{q} - \Sigma_F)_\mu \gamma^\mu + \tilde{\Sigma}_I I}
  {-(\tilde{q}-\Sigma_F)^2 - \tilde{\Sigma}_I^2}
  \right)
  \notag \\ &=
  \log \left[ (\tilde{q} - \Sigma_F)^2 + \tilde{\Sigma}_I^2 \right]
  + \frac{1}{2}
  \frac{(g - 2f^2) q_s^2 - 2 f \sigma q_s}{
  (\tilde{q}-\Sigma_F)^2 + \tilde{\Sigma}_I^2}
  \ed
\end{align}
Notice that
\begin{equation}
  (\tilde{q}-\Sigma_F)^2 + \tilde{\Sigma}_I^2 =
  \tilde{q}_3^2 + (q-\Sigma_F)_s^2 + \tilde{\Sigma}_I^2
  = \tilde{q}_3^2 + q_s^2(1 - g + f^2)
  = \tilde{q}^2 + \beta^{-2} \mu_F^2 \ec
\end{equation}
where we used \eqref{gfc}. The free energy can therefore be written as
\begin{align}
  \beta F_F &= - N V_2 \int_{-1/2}^{1/2} du
  \int \! \frac{d^2q}{(2\pi)^2} \sum_{n=-\infty}^{\infty}
  \left[
  \log \left( \tilde{q}^2 + \beta^{-2} \mu_F^2 \right)
  + \frac{1}{2}
  \frac{(g - 2f^2) q_s^2 - 2 f \sigma q_s}{
  \tilde{q}^2 + \beta^{-2} \mu_F^2}
  \right] \ed
  \label{F-fr-2}
\end{align}
We first consider the log term. The sum over momentum in the compact direction diverges, and we regulate it as before by subtracting the contribution from a massive fermion with mass $\beta^{-1} M$:
\begin{align}
 \int \! \frac{d^2q}{(2\pi)^2}
  & \sum_{n=-\infty}^{\infty}
  \log \left[ \frac{\tilde{q}^2 + \beta^{-2} \mu_F^2}
  {\tilde{q}^2 + \beta^{-2} M^2} \right]
  =
  \int \! \frac{d^2q}{(2\pi)^2}
  \log \left( \prod_{n=-\infty}^\infty \left[
  1 + \frac{(\beta q_s)^2 + \mu_F^2}{\left( 2\pi(n+\frac{1}{2}-|\lambda|u) \right)^2}
  \right] \right)
  - (\mu_F \to M)\nonumber
  \\ &=
  \frac{1}{2\pi\beta^2}
  \int_0^{\Lambda'} \! dx \, x
  \log \left[ 2 \cosh (\sqrt{x^2 + \mu_F^2}) + 2 \cos (2\pi |\lambda|u) \right]
  - (\mu_F \to M) \ed
  \label{tr-fr-1}
\end{align}
The remaining radial integral is the same as that of \eqref{F-logterm-sc} if we replace $\mu_B$ by $\mu_F$ and flip the sign of $\cos(2\pi |\lambda|u)$. After subtracting the divergences as we did in the bosonic case, we find the result
\begin{align}
  \frac{1}{2\pi \beta^2}
  \left[
  - \frac{\mu_F^3}{3}
  + \int_{\mu_F}^\infty \! dy \, y
  \log \left(
  1 + e^{-2y} + 2 e^{-y} \cos(2\pi |\lambda|u)
  \right)
  \right] \ed
\end{align}
We now compute the integral $\int_{-1/2}^{1/2} du$ of this expression. The result can be obtained from \eqref{F-term1-sc} by analytically continuing $y \to y + \pi i$, and we find
\begin{align}
  - \frac{1}{2\pi \beta^2}
  \left\{
  \frac{\mu_F^3}{3}
  + \frac{1}{\pi i |\lambda|}
  \int_{\mu_F}^\infty \! dy \, y
  \left[
  \Li_2 (-e^{-y+\pi i |\lambda|}) - \Li_2 (-e^{-y-\pi i |\lambda|})
  \right]
  \right\}
  \ed
  \label{F-term1-fr}
\end{align}
Let us now consider the second term in \eqref{F-fr-2}. The sum was already computed in \eqref{G-fr}, and we find that
\begin{align}
  &\quad
  \int_{-1/2}^{1/2} du
  \int \! \frac{d^2q}{(2\pi)^2} \sum_{n=-\infty}^{\infty}
  \frac{1}{2}
  \frac{(g - 2f^2) q_s^2 - 2 f \sigma q_s}{
  \tilde{q}^2 + \beta^{-2} \mu_F^2}\nonumber
  \\ &=
  \frac{1}{16 \pi \beta^2}
  \int_{-1/2}^{1/2} du
  \int_0^{\Lambda'} \! dx \, x
  \frac{(g - 2f^2) x^2 - 2 f \tilde{\sigma} x}{\sqrt{x^2 + \mu_F^2}}
  \left[
  \tanh \left( \frac{\sqrt{x^2 + \mu_F^2}+2\pi i |\lambda|u}{2} \right) + \cc
  \right]\nonumber
  \\ &=
  \frac{1}{8\pi^2 i |\lambda| \beta^2}
  \int_0^{\Lambda'} \! dx \, x
  \frac{\mu_F^2 + f(x)^2 x^2 + 2 f(x) \tilde{\sigma} x}{\sqrt{x^2 + \mu_F^2}}
  \left(
  \log \left[ \cosh \left( \frac{\sqrt{x^2 + \mu_F^2} - \pi i |\lambda|}{2} \right) \right]
  - \cc
  \right) \ed
  \label{term2-fr}
\end{align}
Notice that from \eqref{f-fer} we have
\begin{align}
  \pi i \, \frac{\p (x f(x))}{\p x} &=
  \frac{x}{\sqrt{x^2+\mu_F^2}} \left(
  \log \left[ \cosh \left( \frac{\sqrt{x^2 + \mu_F^2} - \pi i \lambda}{2} \right) \right]
  - \cc
  \right)
  \ed
\end{align}
Making the replacement $v = x f(x)$ in \eqref{term2-fr} therefore allows us to solve the integral, with the result
\begin{align}
  \int_{-1/2}^{1/2} du
  \int \! \frac{d^2q}{(2\pi)^2} \sum_{n=-\infty}^{\infty}
  \frac{1}{2}
  \frac{(g - 2f^2) q_s^2 - 2 f \sigma q_s}{
  \tilde{q}^2 + \beta^{-2} \mu_F^2}
  &=
  \frac{1}{2\pi \beta^2} \left(
  \pm \frac{\mu_F^3}{3 \lambda} - \frac{\mu_F^2 \tilde{\sigma}}{2\lambda} + \frac{\tilde{\sigma}^3}{6\lambda}
  \right) \ed
  \label{F-term2-fr}
\end{align}
Combining \eqref{F-term1-fr} and \eqref{F-term2-fr} into \eqref{F-fr-2}, we find that the free energy is given by
\begin{align}
  \beta F_F &=
  \frac{N V_2}{2\pi \beta^2} \left\{
  \frac{\mu_F^3}{3} \left( 1 \mp \frac{1}{\lambda} \right)
  + \frac{\tilde{\sigma} \mu_F^2}{2\lambda} - \frac{\tilde{\sigma}^3}{6\lambda}
  + \frac{1}{\pi i \lambda}
  \int_{\mu_F}^\infty \! dy \, y
  \left[
  \Li_2 \left( - e^{-y+\pi i \lambda} \right) - \cc
  \right]
  \right\} \ed
  \label{F-nmass-nc-fer}
\end{align}
The choices of signs here and in the previous equation depend on the sign for which there is
a solution to \eqref{mass-nc-fer}.
The thermal mass and free energy in the regular theory are obtained by setting $\tilde{\sigma}=0$ in \eqref{mass-nc-fer} and in \eqref{F-nmass-nc-fer}. We can always rewrite \eqref{F-nmass-nc-fer} as
\begin{align}
  \beta F_F &=
  \frac{N V_2}{2\pi \beta^2} \left\{
  \frac{\tilde{\sigma} \mu_F^2}{6\lambda} - \frac{\tilde{\sigma}^3}{6\lambda}
  + \frac{1}{\pi i \lambda} \left[
  \frac{\mu_F^2}{3} \Li_2 \left(- e^{-\mu_F + \pi i \lambda} \right) +
  \int_{\mu_F}^\infty \! dy \, y
  \Li_2 \left( - e^{-y+\pi i \lambda} \right) - \cc
  \right]
  \right\} \ed
  \label{F-nmass-nc-fern}
\end{align}

We can now perform a first test of the duality between the fermionic and bosonic theories. The duality implies that the expressions we found for $\tilde{\sigma}=0$ should map to the analogous expressions \eqref{mass-crit-sc} and \eqref{F-crit-sc} of the critical bosonic theory. One can verify that this is indeed the case, using the duality map
\begin{align}
  |\lambda_b| + |\lambda_f| = 1 \ecq
  \frac{N_b}{|\lambda_b|} = \frac{N_f}{|\lambda_f|} \ecq
  \sign(\lambda_b) = -\sign(\lambda_f) \ec
  \label{duality}
\end{align}
where $N_b$, $\lambda_b$ ($N_f$, $\lambda_f$) are the parameters of the critical bosonic (regular fermionic) theory. In particular, we see that in the $\lambda\rightarrow 1$ limit the thermal mass $\mu_F$ of the regular fermion theory goes to the thermal mass (\ref{On-crit-mass}) of the ordinary critical boson, and the fermion free energy goes to
\begin{equation}
\beta F_F \stackrel{\lambda\rightarrow 1}{\simeq}-N V_2 T^2 (1-\lambda) \frac{4}{5}\frac{\zeta(3)}{\pi} \simeq - k_{\YM} V_2 T^2 \frac{4}{5}\frac{\zeta(3)}{\pi} \ec
\end{equation}
which is the free energy of the bosonic critical $U(k_{\YM})$ model \cite{Sachdev:1993pr}.

\subsection{The critical theory}

Consider next the critical fermionic theory. Like the regular bosonic theory, this has (at large $N$) an exactly marginal triple-trace deformation, that we should add also on the fermionic side in order to map to the bosonic theory with general values of $\lambda_6$. At large $N$, this theory can be defined by deforming the regular theory by $\delta \mathcal{L}_{\text{fermion}} = \sigma\bar{\psi}\psi + \frac{N}{3!}\lambda_6^f \sigma^3$, computing the effective action for $\sigma$, and finally computing the path integral over $\sigma$ (which at large $N$ is equivalent to extremizing the effective action with respect to $\sigma$). The free energy \eqref{F-nmass-nc-fer} is the effective action for $\sigma$ in the thermal theory, and turning on the $\lambda_6^f$ deformation corresponds to adding a term
\begin{align}
  \delta (\beta F_F) = \frac{N V_2}{6\beta^2} \lambda_6^f \tilde{\sigma}^3 \ed
\end{align}
Extremizing $\beta F_F$ with respect to $\sigma$, we obtain
\begin{align}
  \tilde{\sigma} = -\sign(\lambda) \mu_F \hat{g} \ecq
  \hat{g} \equiv \frac{1}{\sqrt{1-2\pi\lambda\lambda_6^f}} \ec
\end{align}
where the sign of $\sigma$ was chosen such that there is a solution of the thermal mass equation with $\mu_F>0$ for any $\lambda$ and $\lambda_6^f$ such that $2\pi\lambda\lambda_6^f < 1$. The dominant solution (with the lowest free energy) has the sign in \eqref{mass-nc-fer} equal to the sign of $\lambda$. The free energy in the critical fermionic theory is then given by (with an `$f$' subscript on fermionic couplings)
\begin{gather}
  \beta F_F^{\mathrm{crit.}} = -\frac{N V_2}{2\pi \beta^2 |\lambda_f|}
  \left[
  \frac{\mu_{F,c}^3}{3}
  \left( 1 - |\lambda_f| + |\hat{g}_f| \right)
  + \frac{1}{\pi i} \int_{\mu_{F,c}}^{\infty} dy y
  \left( \Li_2(-e^{-y-\pi i |\lambda_f|}) - \cc\right)
  \right] \ec
  \label{F-crit-fer}
\end{gather}
where $\mu_{F,c}$ obeys
\begin{align}
  \left( 1 - |\lambda_f| + |\hat{g}_f| \right)
  \mu_{F,c} =
  \frac{1}{\pi i} \left[
  \Li_2\left( -e^{-\mu_{F,c}-\pi i |\lambda_f|} \right) - \cc
  \right] \ed
  \label{mass-crit-fer}
\end{align}
For $\lambda=\lambda_6=0$ the free energy of the critical fermion theory is the same (at large $N$) as that of the free fermion theory (see \cite{Petkou:1998wd,Christiansen:1999uv} and references therein), but this is not true for finite couplings.

Let us now compare this result with the free energy of the regular bosonic theory. Up to normalization, $\sigma$ is dual to the scalar operator $J_0=\phi^\dagger \phi$ in the regular bosonic theory, and the $\sigma^3$ interaction is dual to the bosonic triple-trace deformation $\delta \mathcal{L}_{\text{boson}} = \frac{\lambda_6^b}{3! N_b^2} (J_0)^3$. By matching the $3$-point functions $\langle \sigma \sigma \sigma \rangle$ in the critical-fermionic theory with $\langle J_0 J_0 J_0 \rangle$ in the regular bosonic theory, the mapping of the triple-trace couplings in the two theories is \cite{fermion_corrs}
\begin{align}
\lambda_6^b = 8\pi^2(1-|\lambda_f|)^2\left( 3 - 8\pi \lambda_f \lambda_6^f \right) \ed
\label{lambda6map}
\end{align}
Under the duality map $\eqref{duality}$ and \eqref{lambda6map}, we have that $\hat{\lambda}_b^2 = \lambda_b^2 + \frac{\lambda_6^b}{8\pi^2}$ maps as
\begin{align}
  \frac{|\hat{\lambda}_b|}{|\lambda_b|}
  = \frac{2}{|\hat{g}_f|} \ed
\end{align}
One can now easily check that under this map the free energy \eqref{F-crit-fer} and thermal mass \eqref{mass-crit-fer} of the critical fermionic theory map to those of the regular bosonic theory, given by \eqref{F-nc-sc} and \eqref{mass-nc-sc}. This is true for any value of $\lambda$ and $\lambda_6$ on both sides, with the constraint
\begin{align}
  \lambda_6^f < \frac{1}{2\pi\lambda_f}
  \quad \Longleftrightarrow \quad
  \lambda_6^b > -8\pi^2 \lambda_b^2 \ed
\end{align}

\section{Theories with Bosons and Fermions}
\label{bos-fer}

In this section we generalize the computation of the free energy to massless theories with both fermions and scalars, with generic couplings. As a special case we will consider the theory with $\cN=2$ supersymmetry (for a review, see \cite{Gaiotto:2007qi}). As a test of the computation, we will verify the Seiberg-like duality in this theory that was found in \cite{Benini:2011mf} (generalizing the duality for non-chiral theories found in \cite{Giveon:2008zn}).

The theory we consider includes a single complex scalar and a single Dirac fermion, both in the fundamental representation of $U(N)$, coupled to a gauge field with Chern-Simons interactions. It is defined by the action
\begin{align}
  S &= S_{\CS} + S_{\scalar} + S_{\fermion} +
  \int \! d^3x \, \frac{\lambda_4}{N} (\bar{\psi}\psi) (\phi^\dagger \phi) \ec
\end{align}
where the first three terms are defined in \eqref{SCS}, \eqref{Sscalar}, and \eqref{Sfermion}, and we take the fermion and scalar to be massless for simplicity ($\sigma=0$).\footnote{\label{2tr-couplings} One can write down additional marginal interactions of the form $(\bar{\psi} \phi) (\phi^\dagger \psi)$ and $((\bar{\psi} \phi) (\bar{\psi}\phi) + \cc)$, but they will not affect the free energy in the large $N$ limit.}

\subsection{Exact propagators}

\begin{figure}
  \centering
  \includegraphics[width=0.8\textwidth]{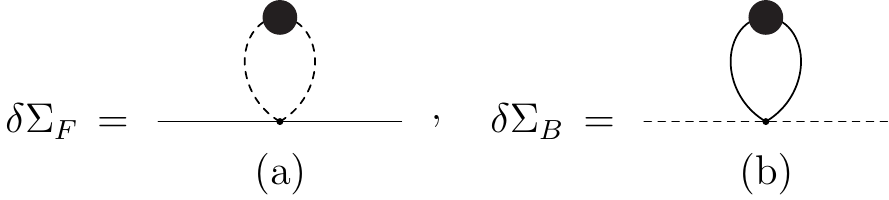}
  \caption{Additional diagrams due to the $\lambda_4$ vertex, in the bootstrap
  equations for the exact propagators.}
  \label{fig:sc-fer-diagrams}
\end{figure}

The scalar and fermion propagators are given in \eqref{exact-prop-sc} and \eqref{exact-prop-fer}, respectively. There is one additional diagram for each bootstrap equation, shown in figure \ref{fig:sc-fer-diagrams}.
These diagrams are given by
\begin{align}
  \mathrm{(a)} &= - \lambda_4 \delta_{ji} I
  \int_{-1/2}^{1/2} du \,
  \int \! \frac{d^2q}{(2\pi)^2} \, \frac{1}{\beta} \sum_n
  \frac{1}{\tilde{q}^2 - \Sigma_B(q)}
  \ec \label{diag-a}
  \\
  \mathrm{(b)} &=
  \lambda_4 \delta_{ji} \int_{-1/2}^{1/2} du \,
  \int \! \frac{d^2q}{(2\pi)^2} \, \frac{1}{\beta} \sum_n
  \tr_f \left(
  \frac{1}{i \tilde{q}_\mu \gamma^\mu - \Sigma_F(q)}
  \right)
  \ed \label{diag-b}
\end{align}
Using the same arguments as in the separate bosonic and fermionic theories, we find again that $\Sigma_- = \Sigma_3 = 0$, and that
\begin{align}
  \Sigma_B(p) &= -\beta^{-2} \mu_B^2(\lambda) \ec \\
  \Sigma_I(p) &= f(y,\lambda) p_s \ec \\
  \Sigma_+(p) &= g(y,\lambda) p_+ \ec
\end{align}
where
\begin{align}
  g - f^2 = - \frac{\mu_F^2(\lambda)}{y^2} \ed
\end{align}
The diagrams can be computed using the same techniques we used above. Let us define
\begin{align}
  h_B &\equiv |\lambda| \mu_B + \frac{1}{\pi i}
  \left[ \Li_2(e^{-\mu_B - \pi i |\lambda|}) - \cc \right]
  \ec \label{hB} \\
  h_F &\equiv |\lambda| \mu_F + \frac{1}{\pi i}
  \left[ \Li_2(-e^{-\mu_F - \pi i |\lambda|}) - \cc \right]
  \ed
  \label{hF}
\end{align}
The integrals in \eqref{diag-a} and \eqref{diag-b} then evaluate to
\begin{align}
  \int \! du \,
  \int \! \frac{d^2q}{(2\pi)^2} \, \frac{1}{\beta} \sum_n
  \frac{1}{\tilde{q}^2 - \Sigma_B(q)}
  &=
  - \frac{h_B}{4\pi\beta|\lambda|}  \ec
  \label{int-sc-loop}
\end{align}
\begin{align}
  \quad \int \! du \,
  & \int \! \frac{d^2q}{(2\pi)^2} \, \frac{1}{\beta} \sum_n
  \tr_f \left(
  \frac{1}{i \tilde{q}_\mu \gamma^\mu - \Sigma_F(q)}
  \right) =
  \notag \\ &\quad
  - \frac{i}{2\pi^2\beta^2 |\lambda|}
  \int_0^{\Lambda'} dx \frac{x^2 f(x)}{\sqrt{x^2+\mu_F^2}}
  \left[
  \log \left( \cosh \left( \frac{\sqrt{x^2+\mu_F^2} - \pi i |\lambda|}{2} \right) \right)
  - \cc
  \right] =
  \notag \\ &\quad
  - \frac{1}{4\pi\beta^2\lambda}
  h_F \left( h_F - \frac{\lambda_4}{2\pi\lambda} h_B \right)
  \ed
  \label{int-fer-loop}
\end{align}
To compute the last integral in \eqref{int-fer-loop} we first determined $f(y)$, which using the results of the previous section and \eqref{int-sc-loop} evaluates to
\begin{align}
  y f(y) &=
  - \lambda \sqrt{y^2 + \mu_F^2}
  + \frac{\sign(\lambda)}{\pi i}
  \left[
  \Li_2 \left( - e^{-\sqrt{y^2 + \mu_F^2} + \pi i |\lambda|} \right)
  - \cc
  \right] + \frac{\lambda_4}{4\pi|\lambda|} h_B \ed
\end{align}


Adding the contributions \eqref{diag-a} and \eqref{diag-b} to the bootstrap equations \eqref{bs-sc} and \eqref{f-fer}, we find the corrected equations for the thermal masses,
\begin{align}
  \mu_B^2 &=
  \left( \frac{\hat{\lambda}}{2{\lambda}} h_B \right)^2
  + \frac{\lambda_4}{4\pi\lambda} h_F
  \left( h_F - \frac{\lambda_4}{2\pi\lambda} h_B \right)
  \ec \label{mu-B}\\
  \mu_F &= h_F - \frac{\lambda_4}{4\pi\lambda} h_B
  \ed \label{mu-F}
\end{align}
In the second line we chose the branch that gives a solution with positive $\mu_F$.

\subsection{Free energy}

The free energy receives separate contributions from the scalars and fermions, as well as contributions from the $\lambda_4$ interaction. It is given by \cite{Jain:2012qi}
\begin{align}
  F = F_B + F_F + F_{BF} \ec
  \label{F-bos-fer}
\end{align}
where $F_B$ is given (in terms of the self-energy) in \eqref{F-sc-1}, $F_F$ in \eqref{F-fr-1}, and
\begin{align}
  \beta F_{BF} &= - \frac{\lambda_4 N V_2 \beta}{6}
  \left[ \int_{-1/2}^{1/2} du \, \int \! \frac{d^3q}{(2\pi)^3} \,
  \frac{1}{\tilde{q}^2 - \Sigma_B}
  \right]
  \left[
  \int_{-1/2}^{1/2} du \, \int \! \frac{d^3p}{(2\pi)^3} \,
  \tr_f \left( \frac{1}{i \tilde{p}_\mu \gamma^\mu - \Sigma_F} \right)
  \right]
  \notag \\ &=
  - \frac{N V_2}{96\pi^2\beta^2} \frac{\lambda_4}{\lambda |\lambda|}
  h_B h_F \left( h_F - \frac{\lambda_4}{2\pi\lambda} h_B \right)
  \ed
\end{align}
In the second line we used the integrals \eqref{int-sc-loop} and
\eqref{int-fer-loop}.

The bosonic contribution $F_B$ to the free energy is given by
\eqref{F-nc-sc-1}, and the fermionic contribution can be computed as in section
\ref{sec:F-fer}. The only difference is in equation \eqref{F-term2-fr}, where
we must account for the corrected thermal mass equation when computing the
integral. We find that
\begin{align}
  \int_{-1/2}^{1/2} du
  \int \! \frac{d^2q}{(2\pi)^2} \frac{1}{\beta} \sum_{n=-\infty}^{\infty}
  \frac{(g - 2f^2) q_s^2}{
  \tilde{q}^2 + \beta^{-2} \mu_F^2}
  &=
  \frac{1}{8\pi|\lambda|\beta^3}
  \left\{
  \frac{1}{3} \left( \frac{\lambda_4}{4\pi\lambda} h_B \right)^3
  + \frac{1}{3} \left( h_F - \frac{\lambda_4}{4\pi\lambda} h_B \right)^3
  + \mu_F^2 h_F
  \right\}
  \ed
\end{align}
The resulting fermionic contribution to the free energy is
\begin{align}
  \beta F_F = \frac{V_2 N}{2\pi\beta^2|\lambda|}
  &\left\{
  \frac{|\lambda| \mu_F^3}{3}
  - \frac{\mu_F^3}{12}
  - \frac{\mu_F^2 h_F}{4}
  - \frac{1}{12} \left( \frac{\lambda_4}{4\pi\lambda} h_B \right)^3
  \right. \notag \\ &\left. \quad
  + \frac{1}{\pi i} \int_{\mu_F}^\infty \! dy \, y
  \left[ \Li_2(-e^{-y+\pi i|\lambda|}) - \cc \right]
  \right\}
  \ed
\end{align}

Combining all the contributions to the free energy and using \eqref{hB}, \eqref{hF} and \eqref{mu-F} to simplify the expression, we obtain
\begin{align}
\beta F &= -\frac{N V_2}{2 \pi^2 i \beta^2 \lambda} \left\{ \frac{\mu_F^2}{3}\left[ \Li_2(-e^{-\mu_F-\pi i \lambda}) - \cc \right] - \frac{\mu_B^2}{3} \left[ \Li_2(e^{-\mu_B-\pi i \lambda}) - \cc \right] \right. \notag\\
    &\quad \left. + \int_{\mu_F}^\infty \! dy \, y
 \left[ \Li_2(-e^{-y-\pi i \lambda}) - \cc \right] - \int_{\mu_B}^\infty \! dy \, y
 \left[ \Li_2(e^{-y-\pi i \lambda}) - \cc \right] \right\} \ed
\end{align}

\subsection{The supersymmetric case}

The theory with $\cN=2$ supersymmetry is conjectured to be self-dual
\cite{Benini:2011mf}, with the transformation at large $N$ given by $N \to |k|
- N$ and $k \to -k$. In this section we verify that the free energy is invariant under this duality.

\begin{figure}
  \centering
  \includegraphics[width=0.6\textwidth]{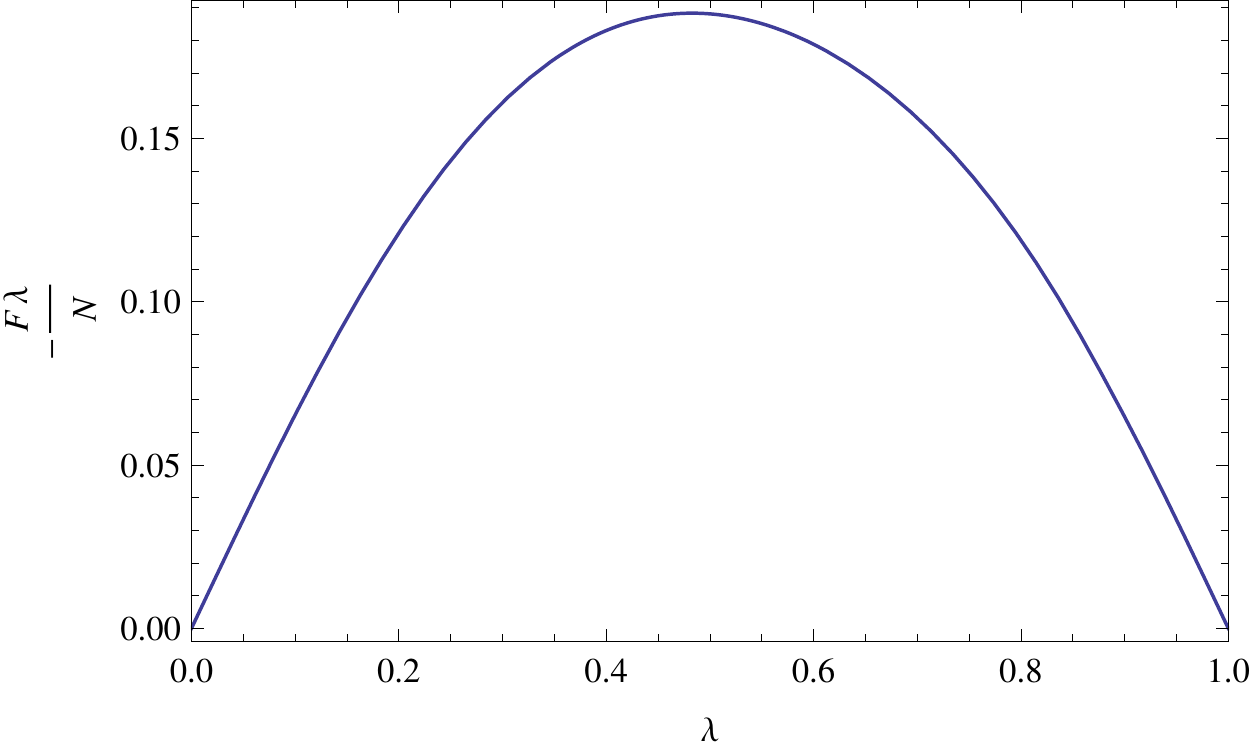}
  \caption{Free energy (times $-\lambda$) in the $\cN=2$ theory, in arbitrary
  normalization.}
  \label{fig:F-susy}
\end{figure}

In this theory, the couplings are given in terms of
$\lambda$ by\footnote{  The supersymmetric theory also contains other couplings which do
 not contribute at large $N$, as mentioned in footnote \ref{2tr-couplings}. }
  \cite{Jain:2012qi}
\begin{align}
  \lambda_4 = 4 \pi \lambda \ecq
  \lambda_6 = 24 \pi^2 \lambda^2 \ec
\end{align}
so that $\hat{\lambda} = 2|\lambda|$.  The equations for the thermal masses become
\begin{align}
  \mu_B^2 &= \left( h_B - h_F \right)^2 \ecq
  \mu_F = h_F - h_B \ed
\end{align}
The thermal masses are therefore equal,
$\mu \equiv \mu_B = \mu_F$,
with $\mu$ satisfying
\begin{align}
  \mu &= \frac{1}{\pi i} \left[
  \Li_2(e^{-\mu + \pi i |\lambda|}) + \Li_2(-e^{-\mu - \pi i |\lambda|})
  -\cc
  \right] \ed
\end{align}
The free energy is given by
\begin{align}
  \beta F &= \frac{V_2 N}{2\pi\beta^2|\lambda|} \left\{
  - \frac{\mu^3}{3}
  + \frac{1}{\pi i} \int_{\mu}^\infty \! dy \, y
  \left[
  \Li_2(e^{-y-\pi i|\lambda|})
  + \Li_2(-e^{-y+\pi i|\lambda|})
  - \cc
  \right]
  \right\} \ed
\end{align}
Both the thermal mass and the free energy are invariant under the duality, which in terms of $N$ and $|\lambda|$ takes $|\lambda| \to 1 - |\lambda|$ with $N/|\lambda|$ fixed.
The free energy is plotted in figure \ref{fig:F-susy}.
It is straightforward to generalize our results also to theories with different amounts of supersymmetry, including the theories analyzed in \cite{Feinberg:2005nx,Jain:2012qi}.

\section{Summary of Results}
\label{summary-sec}
In this section we collect the main results obtained in this paper.

We have computed the thermal free energy on $\mathbb{R}^2$ for various large $N$ vector models coupled to Chern-Simons gauge fields, working exactly in the `t Hooft coupling $\lambda=\frac{N}{k}$. In particular, we have explicitly tested the conjectured non-supersymmetric dualities (``3d bosonization") relating the regular/critical fermion coupled to Chern-Simons to the critical/regular scalar coupled to Chern-Simons \cite{Giombi:2011kc, Maldacena:2012sf, Aharony:2012nh}. Our calculations closely followed \cite{Giombi:2011kc,Jain:2012qi,unp}, with the difference that we included the effect of the holonomy around the thermal circle, as explained in Section \ref{holonomy-sec}.

\begin{figure}[!ht]
\centering
  \includegraphics[width=0.6\textwidth]{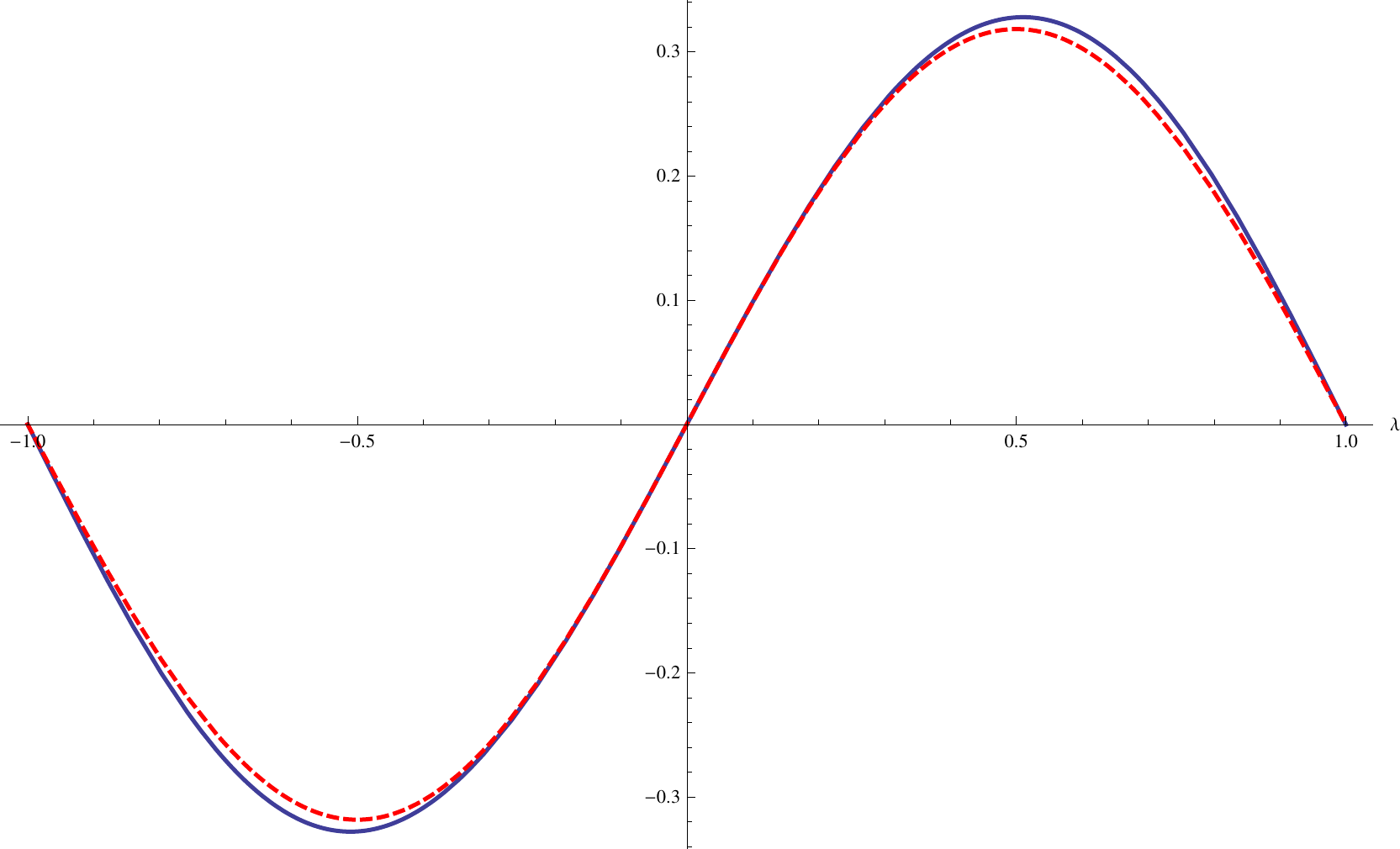}
  \caption{Plot of $(-\lambda) $ times the free energy,  $ (-\lambda F_F)$,  (blue solid line) and
  $\lambda$ times the stress-tensor 2-point function $ \lambda \langle TT\rangle \sim \frac{\sin \pi \lambda}{\pi}$ (red dashed line) for the regular fermion theory.
   They look very similar but they have a different functional
  form.}
  \label{fig:FvsTT}
\end{figure}

In all cases we find that the free energy goes to zero at $\lambda=1$, in accordance with the behavior of the exact $\langle TT\rangle$ correlator \cite{Aharony:2012nh, fermion_corrs}. In fact, since the thermal free energy $F$ and $\langle TT\rangle$ are both a measure of the number of degrees of freedom of the theory, it is interesting to compare them. Amusingly, we find that their behavior as a function of $\lambda$ is rather similar. As an example, figure \ref{fig:FvsTT} shows a plot of $(-\lambda F)$ and $\lambda \langle TT\rangle$ for the regular fermion theory, normalized to agree at weak coupling. Note that it is possible to smoothly continue our results beyond $\lambda=1$.
 In fact, the functions $\lambda F $ and $\lambda \langle TT \rangle $ in the regular fermion theory (and the critical bosonic theory) are
both  periodic under $\lambda \to \lambda + 2$ and odd under $\lambda \to -\lambda$.

Our results are most simply expressed in terms of the following functions
\begin{align}
  \cF(\mu,\lambda) &\equiv
  \frac{1}{2 \pi^2 i} \left[
  \frac{\mu^2}{3} \Li_2( e^{- \mu - \pi i \lambda} )
  + \int_{\mu}^\infty \! dy \, y \,
  \Li_2(e^{-y-\pi i\lambda})
  - \cc \right] \ec
  \\
 h_B(\mu_B , \lambda)  &\equiv \lambda \mu_B + \frac{1}{\pi i}
  \left[ \Li_2(e^{-\mu_B - \pi i \lambda}) - \cc \right] \ec
  \\
  h_F(\mu_F,\lambda)  &\equiv\lambda \mu_F + \frac{1}{\pi i}
  \left[ \Li_2(-e^{-\mu_F - \pi i \lambda}) - \cc \right]
  \ed
\end{align}


\subsection{Regular fermion and critical boson results}

For the regular fermion theory (a massless fermion coupled to CS), the exact planar free energy takes the form
\begin{align} \label{freeenfer}
F_F = -\frac{N V_2 T^3}{\lambda} \cF(\mu_F,\lambda-{\mathrm{sign}}(\lambda)) \ec
\end{align}
where the thermal mass $\mu_F$ is determined by the equation
\begin{align} \label{massfer}
\mu_F = h_F(\mu_F,\lambda)  \ed
\end{align}
Note that here (unlike our discussion above) we allow $\mu_F$ also to be negative, and the equations are still valid in this case.
In fact, we have $h_F(-\mu_F , -\lambda) = - h_F(\mu_F,\lambda)$, and
the free energy is even under $\lambda \to - \lambda$.
For the critical boson coupled to CS, we found
\begin{align}
F_B^{\mathrm{crit.}} = \frac{N V_2 T^3}{\lambda} \cF(\mu_{B,c},\lambda) \ec
\end{align}
where $\mu_{B,c}$ is given by
(again allowing also negative values)
\begin{align}
h_{B}(\mu_{B,c}, \lambda) = 0 \ec
\end{align}

It is now obvious that if the thermal masses map to each other under the duality, so will the free energies.
 Note that ${N \over \lambda } =    k_{\YM} + N \mathrm{sign}(k_{\YM})$
 changes sign under the duality
 It is easy to see that the thermal masses are indeed related by $|\mu_F(\lambda)|=|\mu_{B,c}(\lambda-\sign(\lambda))|$, in agreement with the expected duality.\footnote{Note that the function ${\cal F}(\mu,\lambda)$ is even in $\mu$, so the thermal free energy is insensitive to the sign of $\mu$.} The inclusion of the non-trivial holonomy was crucial in order to obtain this.

Our theories have a global $U(1)$ symmetry, which acts on the scalars/fermions in a natural way (the
equations of motion of the gauge field identify this symmetry with the topological symmetry of the three dimensional $U(1)$ gauge field). It is simple to generalize our analysis by adding a chemical potential for this gauge field, as done (without taking the holonomy into account)
for the regular fermion theory in \cite{Yokoyama:2012fa}.
In the bosonic theories we find that the effect of adding a chemical potential $\mu_{U(1)}$ is simply to replace every
$\Li_2(e^{x})$ appearing in our analysis by $\frac{1}{2} (\Li_2(e^{x+\mu_{U(1)}}) + \Li_2(e^{x-\mu_{U(1)}}))$. The chemical potential maps to itself under the duality.

 For comparison to results that
  previously appeared in the literature, one may also look at the perturbative expansion of these results around $\lambda=0$. The first few orders of the thermal mass and free energy for the regular fermion theory read
\begin{align}
&\mu_F = 2\log (2) \,\lambda +\left(\log^2 (2) -\frac{\pi^2}{12}\right)\lambda^3+{\cal O}(\lambda^5) \ec \nonumber \\
&F_F = -NV_2 T^3 \left[\frac{3\zeta(3)}{4\pi}-\frac{\pi^2\log (2)+4 \log^3\! (2)}{6\pi}\lambda^2+{\cal O}(\lambda^4)\right] \ed
\end{align}
One may verify that these differ from the results in \cite{Giombi:2011kc} already at order $\lambda^2$. Analogously, one may derive the perturbative expansion of the critical boson free energy. Since the thermal mass $\mu_{B,c}(\lambda=0)=2\log \left(\frac{\sqrt{5}+1}{2}\right)$ is non-zero, the perturbative expansion is analytic. By the duality map, this also means that the expansion of the fermion free energy (and thermal mass) close to $\lambda=1$ is analytic, unlike the result found in \cite{Giombi:2011kc}.

\subsection{Regular boson and critical fermion results}

In this case, the situation is slightly more complicated due to the possibility of adding a sextic coupling $\frac{\lambda_6}{6N^2} (J_0)^3$, where $J_0$ is the scalar/fermion bilinear. In the large $N$ limit, this coupling is exactly marginal \cite{Aharony:2011jz} and we have computed the free energies keeping $\lambda_6$ as a free parameter. The result for the regular boson turns out to be
\begin{align}
F_B = \frac{N V_2 T^3}{\lambda} \cF(\mu_B,\lambda) \ec
\end{align}
where the thermal mass is determined by
\begin{align}
\label{sum-mub}
-\frac{2\lambda}{\hat{\lambda}} \mu_B = h_B(\mu_B, \lambda)
 \ec
\end{align}
with $\hat \lambda^2 = \lambda^2+\frac{\lambda_6}{8\pi^2}$.

For the critical fermion theory, we found that the free energy is given by
\begin{align}
F_F^{\mathrm{crit.}} = -\frac{N V_2T^3}{\lambda} \cF(\mu_{F,c},\lambda-{\mathrm{sign}}(\lambda)) \ec
\end{align}
and the thermal mass is given by
\begin{align}
\label{sum-muf}
(1+\hat{g})\mu_{F,c} = h_{F}(\mu_{F,c},\lambda) \ec
\end{align}
where $\hat{g} = 1/\sqrt{1-2\pi\lambda \lambda_6^f}$.

In this case, the duality requires a non-trivial map between the $\lambda_6$ couplings, given in equation (\ref{lambda6map}), which can be derived independently by matching the scalar three-point functions in the two theories \cite{fermion_corrs}. For example, the critical fermionic theory with $\lambda_6^f=0$ ($\hat{g}=1$) maps under $\lambda \leftrightarrow \lambda-{\mathrm{sign}}(\lambda)$ to the regular bosonic theory with $\lambda_6=24\pi^2\lambda^2$ ($\hat\lambda=2|\lambda|$), as can be easily seen from the expressions above.

\subsection{Theories with bosons and fermions }

Let us now summarize the results for the theory with one scalar and one fermion, discussed in section \ref{bos-fer}. This theory contains a boson-fermion coupling $\frac{\lambda_4}{N} (\bar{\psi}\psi)(\phi^{\dag}\phi)$ in addition to the scalar sextic coupling $\frac{\lambda_6}{6N^2}(\phi^{\dag}\phi)^3$. We found that the free energy in this theory is
\begin{align}
F = \frac{ N V_2 T^3}{\lambda} \left( \cF(\mu_B,\lambda) - \cF(\mu_F,\lambda-\mathrm{sign}(\lambda)) \right) \ec
\label{bos-fer-F}
\end{align}
where
\begin{align}
\mu_B^2 &= \left( \frac{\hat{\lambda}}{2\lambda} h_B \right)^2 + \frac{\lambda_4}{4\pi\lambda} h_F \left( h_F - \frac{\lambda_4}{2\pi\lambda} h_B \right) \ec \\
\mu_F &= h_F - \frac{\lambda_4}{4\pi\lambda} h_B \ec
\end{align}
and $\hat{\lambda}$ is defined as in the regular boson theory.

The $\cN=2$ supersymmetric Chern-Simons theory with one chiral superfield is a particular case of the above theory, obtained by setting $\hat{\lambda}=2|\lambda|$ and $\lambda_4 = 4\pi\lambda$. In this case we obtain
\begin{align}
\mu = \mu_F = \mu_B = h_F - h_B \ed
\end{align}
One finds that $\mu$ is self-dual under the level-rank duality transformation, consistent with the expected Seiberg-like duality of this theory \cite{Benini:2011mf}.

One would expect that the theory with one scalar and one fermion is self-dual (at infinite $N$) also away from the $\cN=2$ supersymmetric point, for general values of $\lambda_4$ and $\lambda_6$, since we can deform the two theories away from the supersymmetric point, and the deformation is exactly marginal for infinite $N$ (but not for finite $N$). Let us now describe how this duality works in the large $N$ limit.

The single-trace operators of this theory consist of the operators of the regular boson theory, which we will denote by $J_s^b$ ($s=0,1,\cdots$), and those of the regular fermion theory, denoted by $J_s^f$. In addition, there is a tower of half-integer spin operators obtained by contracting the scalar with the fermion.

It is easy to generalize the computations of planar correlators of integer spin operators, performed in \cite{Aharony:2012nh,fermion_corrs}, to this theory. One finds that the results are consistent with conformal invariance for any $\lambda_4$ and $\lambda_6$. In particular, by computing the exact planar 2-point functions of the operators $(\phi^{\dag}\phi)(\bar{\psi}\psi)$ and $(\phi^{\dag}\phi)^3$ we find that these operators have dimension $3+O(1/N)$, consistent with the expectation that these deformations are exactly marginal in the planar limit\footnote{This fact is also consistent with the perturbative computations of the beta-functions, done in \cite{Avdeev:1992jt}.}.

The planar 3-point functions of the operators $J_s^b$ or $J_s^f$ with $s \ge 1$ are independent of $\lambda_4$ and $\lambda_6$, giving the same results as in the regular boson or fermion theories. Therefore the correlators of the currents $J_s^b$ map to those of $J_s^f$ under the level-rank transformation. The $\lambda_4$ deformation affects planar 3-point functions with insertions of the scalar operators $J_0^b$ and $J_0^f$, while $\lambda_6$ affects only the 3-point function of $J_0^b$.\footnote{Note that $J_0^b$ and $J_0^f$ have different dimensions in this theory ($1+O(1/N)$ and $2+O(1/N)$ respectively), so they must be self-dual. In particular we find that $J_0^b\rightarrow -\lambda\lambda_4 J_0^b$, up to some positive multiplicative factor.} By computing 3-point functions with insertions of $J_0^b$ and $J_0^f$ we find the duality map of $\lambda_4$ and $\lambda_6$ to be
\begin{align}
x_4 \equiv \frac{\lambda_4}{4\pi\lambda} &\rightarrow \frac{1}{x_4} \ec \\
\frac{\hat{\lambda}^2}{4\lambda^2} &\rightarrow \frac{1}{x_4^3}\left( 1 + x_4^3 - \frac{\hat{\lambda}^2}{4\lambda^2} \right) \ed
\end{align}
Note that the above transformations are consistent with the self-duality of the $\cN=2$ theory in which $x_4=\frac{\hat{\lambda}^2}{4\lambda^2}=1$.

One can now verify that the free energy \eqref{bos-fer-F} maps to itself under the above transformations for any values of $\lambda_4$ and $\lambda_6$, consistent with the expected self-duality of this theory. Note that to complete the duality map one has to determine also the mapping of the half-integer spin operators, and include the effects of the additional double-trace deformations of this theory, constructed from the spin $1/2$ operators $(\bar{\psi}\phi)$ and $(\phi^{\dag}\psi)$ (see footnote \ref{2tr-couplings}). While the free energy does not depend on those deformations, correlators of the spin $1/2$ operators will. Moreover, one has to prove that all the above deformations are exactly marginal more carefully. We leave a more detailed analysis of these issues to future work.

\subsection{Massive theories}

While in this paper we have concentrated on the conformal theories, we should note that our results also allow us to readily obtain the thermal free energy for massive fermions or scalars coupled to a Chern-Simons gauge field.
Studying these theories in detail is beyond the scope of this paper, but let us make a few
brief comments about them.

For example, the regular fermion free energy derived in \nref{F-nmass-nc-fer}, with $\mu_F$ obeying
 \nref{mass-nc-fer}, describes a massive fermion theory, where the mass parameter appearing in the Lagrangian is  $ m_F = \tilde \sigma/\beta$.
 For convenience we rewrite here the two equations defining $\tilde m_f = \beta m_f$,
 \begin{align}
    {\mathrm{sign}}(\lambda) \mu_F  =  \tilde  m_f  + \lambda \mu_F + \frac{1}{\pi i}
  \left[
  \Li_2 \left( - e^{-\mu_F-\pi i \lambda} ) - \cc \right)
  \right] \ec
  \label{mass-nc-fersu}
\end{align}
 \begin{align}
  \beta F_F &=
  \frac{N V_2}{2\pi \beta^2} { 1 \over \lambda }  \left\{
 \frac{\mu_F^3}{3} (\lambda - {\mathrm{sign}}(\lambda))
  + \frac{\tilde{m}_f \mu_F^2}{2 } - \frac{ { \tilde m}_f^3}{6 }
  + \frac{1}{\pi i  }
  \int_{\mu_F}^\infty \! dy \, y
  \left[
  \Li_2 \left( - e^{-y+\pi i \lambda } \right) - \cc
  \right]
  \right\} \ed
  \label{F-nmass-nc-fersu}
 \end{align}
 As above, we allow also negative values of $\mu_F$; there is always a solution for either positive or negative $\mu_F$.

This theory should be related by the duality to the critical boson theory, with a mass deformation $m_b$ proportional to $m_f$. In this theory we continue to get equation \nref{mass-crit-sc-1}, but now we need to add an extra
 term to the free energy in \nref{bs-F-sc} of the form
 $ - { \tilde m_b \tilde \sigma \over 2 } $ inside
 the curly brackets of \nref{bs-F-sc} (where again $\tilde m_b \equiv \beta m_b$).
 The equation for $\tilde \sigma$
 is now modified into
 \begin{align}
  {4 \over \lambda^2} \left(\mu_B^2 - \tilde \sigma\right) = \tilde m_b^2 \ed
 \end{align}
 Substituting this value of $\tilde \sigma$ in \nref{mass-crit-sc-1} and \nref{bs-F-sc}, we obtain (using
 here $\mu_B > 0$)
 \begin{align}
   - \lambda \tilde{m}_b  =
  - \lambda \mu_B - \frac{1}{  \pi i}
  \left[ \Li_2(e^{-\mu_B-\pi i \lambda}) - \Li_2(e^{-\mu_B+\pi i \lambda}) \right] \ec
  \label{mass-crit-sc-su}
\end{align}
 \begin{align}
   \beta F_{B,m_b}^{\mathrm{crit.}} = - \frac{N V_2 T^2}{2\pi} \left\{
  \frac{\mu_B^3}{3}
  + { \lambda^2 \tilde{m}_b^3 \over 6 } - { \tilde{m}_b \mu_B^2 \over 2 }
  + \frac{1}{\pi i \lambda}
  \int_{\mu_B}^\infty \! dy \, y
  \left[
  \Li_2 (e^{-y+\pi i \lambda}) - \Li_2 (e^{-y-\pi i \lambda })
  \right]
  \right\}
  \ed
  \label{bs-F-sc-su}
\end{align}
In writing \eqref{bs-F-sc-su} we added to the Lagrangian density a constant term proportional to $N \lambda^2 m_b^3$ (this modifies only the second term, and does not affect the physics), in order
to match with the fermionic theory, using the identifications
\begin{align}
  |\mu_B| = |\mu_F| , ~~~~~~~~~\lambda_b = \lambda_f - \mathrm{sign}(\lambda_f) , ~~~~~~~\lambda_b m_b = -m_f \ed
\end{align}
The last relation between the mass parameters follows (up to a sign) from properly normalizing the
associated spin zero operators, using their two-point functions computed in \cite{Aharony:2012nh,fermion_corrs}.

Actually, this is not precise, since equation \eqref{mass-crit-sc-su} does not always have a solution (recall $\mu_B > 0$). When it does not, it turns out that there is (at least at zero temperature \cite{Aharony:2012nh}) a different configuration in which one of the scalars obtains an expectation value, breaking the $U(N)$ gauge symmetry to $U(N-1)$. The duality implies that analyzing this configuration
(using the expectation value that minimizes
the free energy) should lead to exactly the same equations that we had above, just for negative values of
$\mu_B$, in order to be consistent with the matching with the fermionic theory for these values of the mass. It would be interesting to confirm this.

It is interesting to take the zero temperature limit of the above relations. In the
zero temperature limit $\tilde m$ and $\mu$ become large, so the dilogarithms go to zero if the argument of their exponent is negative (otherwise we can map to this situation, using the identity $\Li_2(x) + \Li_2(1/x) = -\frac{\pi^2}{6} - \frac{1}{2} (\log(-x))^2$). For the two theories described above, we find the simple results (assuming for simplicity $ 0 < \lambda_b < 1 $, $-1 < \lambda_f < 0$)
\begin{align}
    { \mu_B \over \beta}  = -{ \mu_F \over \beta}  = &   m_b= -{   m_f \over (1 + \lambda_f)  } ~,~~~~~~~~~~~~~ \qquad m_{b} > 0, m_f < 0 \ec
   \\
  { \mu_B \over \beta}  = -{ \mu_F \over \beta}  = &
   { \lambda_b   m_b\over (2 - \lambda_b)} = -{   m_f \over (1 - \lambda_f ) }
    ~,~~~~~~~~~~~~~m_{b} < 0, m_f > 0 \ed
  \label{larma}
\end{align}

Note that $\mu_{F,B}/\beta$ is the position of the pole of the dressed fermion/scalar propagator in these theories. This is not a gauge-invariant quantity.
However, at least for weak coupling, we expect that the position of
this pole is close to the actual physical mass of the lightest (charged) excitation. In the massive theory, the low
energy effective description is by a topological pure Chern-Simons theory. We have massive excitations whose worldlines act as Wilson line sources for this Chern-Simons theory,
with vanishing forces between them at leading order in the large $N$ limit.

In the fermionic theory for small $\lambda_f$ the excitation mass is of order $m_f$. However,
it seems strange that when $\lambda_f$ is not small,
we have a significant difference between the two choices
of signs for $m_f$; the low-energy theories in the two cases differ just by shifting $k$ by one (which is negligible in the large $N$ limit). However, we can understand the origin of this difference by thinking
about the critical bosonic theory.
In this
theory we see that for small $\lambda_b$ the physical mass of the excitations
is the same as $m_b$ for $m_b > 0$, but is a much smaller scale of order $\lambda_b m_b$ for negative $m_b$.
As mentioned in \cite{Aharony:2012nh}, this is because for one sign the theory is in
a phase with unbroken $U(N)$, while for the other sign the $U(N)$ symmetry is spontaneously
broken to $U(N-1)$ by a scalar vacuum expectation value.
Since the
$U(N)$ symmetry is gauged, instead of having Nambu-Goldstone bosons, we have light massive
gauge bosons with a mass of order $\lambda_b m_b$ (for $\lambda_b \ll 1$).

There is also another way to obtain massive theories. Instead of a massive deformation, in
some cases we have (for zero temperature in the regular bosonic and critical fermionic theories)
a moduli space along which we can spontaneously
break the conformal symmetry by a vacuum expectation value  for the gauge-invariant scalar operator.
Without the Chern-Simons coupling this possibility was analyzed in \cite{Bardeen:1983rv, Amit:1984ri}. Usually the only solution
to our equations at zero temperature is $\mu=0$. However, in our regular bosonic theories, we see that when $\hat\lambda=2$ (namely $\lambda^2 + \frac{\lambda_6}{8\pi^2} = 4$), there is a solution to the ``gap equation'' \eqref{sum-mub} for the thermal mass for every negative value of $\mu_B$ (in the zero temperature limit). This is the manifestation of the opening of a new moduli space of vacua for this specific value of $\hat\lambda$ in our analysis. Along this moduli space the conformal symmetry is spontaneously broken, and there is a massless Nambu-Goldstone boson. Of course, for finite temperature this moduli space is lifted, and we also do not expect it to be present for finite values of $N$.

\subsection*{Acknowledgments}

We would like to thank S. Banerjee, S. Hellerman, I. Klebanov, J. Maltz, G. Moore,  S. Minwalla, M. Moshe, E. Rabinovici, S. Shenker, J. Sonnenschein and X. Yin for useful discussions and correspondence.
OA is the Samuel Sebba Professorial Chair of Pure and Applied Physics, and he is supported in part by a grant from the Rosa and Emilio Segre Research Award. The work of OA, GG and RY was supported in part by an Israel Science Foundation center for excellence grant, by the German-Israeli Foundation (GIF) for Scientific Research and Development, and by the Minerva foundation with funding from the Federal German Ministry for Education and Research. OA gratefully acknowledges support from an IBM Einstein Fellowship at the Institute for Advanced Study.
The work of JM was supported in part by   U.S.~Department of Energy grant \#DE-FG02-90ER40542.



\begin{thebibliography}{99}


\bibitem{Giombi:2011kc}
  S.~Giombi, S.~Minwalla, S.~Prakash, S.~P.~Trivedi, S.~R.~Wadia and X.~Yin,
  ``Chern-Simons Theory with Vector Fermion Matter,''
  Eur.\ Phys.\ J.\ C {\bf 72}, 2112 (2012)
  [arXiv:1110.4386 [hep-th]].



\bibitem{Aharony:2011jz}
  O.~Aharony, G.~Gur-Ari and R.~Yacoby,
  ``d=3 Bosonic Vector Models Coupled to Chern-Simons Gauge Theories,''
  JHEP {\bf 1203} (2012) 037
  [arXiv:1110.4382 [hep-th]].


\bibitem{Chang:2012kt}
  C.~-M.~Chang, S.~Minwalla, T.~Sharma and X.~Yin,
  ``ABJ Triality: from Higher Spin Fields to Strings,''
  arXiv:1207.4485 [hep-th].


\bibitem{Jain:2012qi}
  S.~Jain, S.~P.~Trivedi, S.~R.~Wadia and S.~Yokoyama,
  ``Supersymmetric Chern-Simons Theories with Vector Matter,''
  arXiv:1207.4750 [hep-th].

\bibitem{Aharony:2012nh}
  O.~Aharony, G.~Gur-Ari and R.~Yacoby,
  ``Correlation Functions of Large N Chern-Simons-Matter Theories and Bosonization in Three Dimensions,''
  arXiv:1207.4593 [hep-th].

\bibitem{fermion_corrs}
  G.~Gur-Ari and R.~Yacoby,
  ``Correlators of Large N Fermionic Chern-Simons Vector Models,''
  arXiv:1211.1866 [hep-th].

\bibitem{Maldacena:2011jn}
  J.~Maldacena and A.~Zhiboedov,
  ``Constraining Conformal Field Theories with A Higher Spin Symmetry,''
  arXiv:1112.1016 [hep-th].

\bibitem{Maldacena:2012sf}
  J.~Maldacena and A.~Zhiboedov,
  ``Constraining conformal field theories with a slightly broken higher spin symmetry,''
  arXiv:1204.3882 [hep-th].

\bibitem{Vasiliev:1992av}
  M.~A.~Vasiliev,
  ``More on equations of motion for interacting massless fields of all spins in (3+1)-dimensions,''
  Phys.\ Lett.\ B {\bf 285}, 225 (1992).

\bibitem{Vasiliev:1999ba}
  M.~A.~Vasiliev,
  ``Higher spin gauge theories: Star product and AdS space,''
  In *Shifman, M.A. (ed.): The many faces of the superworld* 533-610
  [hep-th/9910096].

\bibitem{Giombi:2012ms}
  S.~Giombi and X.~Yin,
  ``The Higher Spin/Vector Model Duality,''
  arXiv:1208.4036 [hep-th].

\bibitem{Klebanov:2002ja}
  I.~R.~Klebanov and A.~M.~Polyakov,
  ``AdS dual of the critical O(N) vector model,''
  Phys.\ Lett.\ B {\bf 550}, 213 (2002)
  [hep-th/0210114].

\bibitem{Sezgin:2003pt}
  E.~Sezgin and P.~Sundell,
  ``Holography in 4D (super) higher spin theories and a test via cubic scalar couplings,''
  JHEP {\bf 0507}, 044 (2005)
  [hep-th/0305040].

\bibitem{Witten:1988hf}
  E.~Witten,
  ``Quantum Field Theory and the Jones Polynomial,''
  Commun.\ Math.\ Phys.\  {\bf 121}, 351 (1989).

\bibitem{Naculich:1990pa}
  S.~G.~Naculich, H.~A.~Riggs and H.~J.~Schnitzer,
  ``Group Level Duality In Wzw Models And Chern-simons Theory,''
  Phys.\ Lett.\ B {\bf 246}, 417 (1990).

\bibitem{Camperi:1990dk}
  M.~Camperi, F.~Levstein and G.~Zemba,
  ``The Large N Limit Of Chern-simons Gauge Theory,''
  Phys.\ Lett.\ B {\bf 247}, 549 (1990).

\bibitem{Mlawer:1990uv}
  E.~J.~Mlawer, S.~G.~Naculich, H.~A.~Riggs and H.~J.~Schnitzer,
  ``Group level duality of WZW fusion coefficients and Chern-Simons link observables,''
  Nucl.\ Phys.\ B {\bf 352}, 863 (1991).

\bibitem{Naculich:2007nc}
  S.~G.~Naculich and H.~J.~Schnitzer,
  ``Level-rank duality of the U(N) WZW model, Chern-Simons theory, and 2-D qYM theory,''
  JHEP {\bf 0706} (2007) 023
  [hep-th/0703089].

\bibitem{Kapustin:2010mh}
  A.~Kapustin, B.~Willett and I.~Yaakov,
  ``Tests of Seiberg-like Duality in Three Dimensions,''
  arXiv:1012.4021 [hep-th].

\bibitem{unp}
  O.~Aharony, G.~Gur-Ari and R.~Yacoby, unpublished.

\bibitem{Sundborg:1999ue}
  B.~Sundborg,
  ``The Hagedorn transition, deconfinement and N=4 SYM theory,''
  Nucl.\ Phys.\ B {\bf 573} (2000) 349
  [hep-th/9908001].

\bibitem{Aharony:2003sx}
  O.~Aharony, J.~Marsano, S.~Minwalla, K.~Papadodimas and M.~Van Raamsdonk,
  ``The Hagedorn - deconfinement phase transition in weakly coupled large N gauge theories,''
  Adv.\ Theor.\ Math.\ Phys.\  {\bf 8} (2004) 603
  [hep-th/0310285].

\bibitem{Giveon:2008zn}
  A.~Giveon and D.~Kutasov,
  ``Seiberg Duality in Chern-Simons Theory,''
  Nucl.\ Phys.\ B {\bf 812} (2009) 1
  [arXiv:0808.0360 [hep-th]].

\bibitem{Benini:2011mf}
  F.~Benini, C.~Closset and S.~Cremonesi,
  ``Comments on 3d Seiberg-like dualities,''
  JHEP {\bf 1110}, 075 (2011)
  [arXiv:1108.5373 [hep-th]].

\bibitem{Aharony:2008ug}
  O.~Aharony, O.~Bergman, D.~L.~Jafferis and J.~Maldacena,
  ``N=6 superconformal Chern-Simons-matter theories, M2-branes and their gravity duals,''
  JHEP {\bf 0810} (2008) 091
  [arXiv:0806.1218 [hep-th]].


\bibitem{Banerjee:2012gh}
  S.~Banerjee, S.~Hellerman, J.~Maltz and S.~H.~Shenker,
  ``Light States in Chern-Simons Theory Coupled to Fundamental Matter,''
  arXiv:1207.4195 [hep-th].

\bibitem{Banerjee:2012aj}
  S.~Banerjee, A.~Castro, S.~Hellerman, E.~Hijano, A.~Lepage-Jutier, A.~Maloney and S.~Shenker,
  ``Smoothed Transitions in Higher Spin AdS Gravity,''
  arXiv:1209.5396 [hep-th].

\bibitem{LindeTS}
  A.~D.~Linde,
  ``Infrared Problem in Thermodynamics of the Yang-Mills Gas,''
Phys.\ Lett.\ B {\bf 96}, 289 (1980)..

\bibitem{Svetitsky:1985ye}
  B.~Svetitsky,
  ``Symmetry Aspects of Finite Temperature Confinement Transitions,''
  Phys.\ Rept.\  {\bf 132}, 1 (1986).

\bibitem{Braaten:1994qx}
  E.~Braaten and A.~Nieto,
  ``Asymptotic behavior of the correlator for Polyakov loops,''
  Phys.\ Rev.\ Lett.\  {\bf 74}, 3530 (1995)
  [hep-ph/9410218].

\bibitem{Brezin:1977sv}
  E.~Brezin, C.~Itzykson, G.~Parisi and J.~B.~Zuber,
  ``Planar Diagrams,''
  Commun.\ Math.\ Phys.\  {\bf 59}, 35 (1978).

\bibitem{Douglas:1994ex}
  M.~R.~Douglas,
  ``Chern-Simons-Witten theory as a topological Fermi liquid,''
  hep-th/9403119.

\bibitem{Elitzur:1989nr}
  S.~Elitzur, G.~W.~Moore, A.~Schwimmer and N.~Seiberg,
  ``Remarks on the Canonical Quantization of the Chern-Simons-Witten Theory,''
  Nucl.\ Phys.\ B {\bf 326} (1989) 108.

\bibitem{Blau:1993tv}
  M.~Blau and G.~Thompson,
  ``Derivation of the Verlinde formula from Chern-Simons theory and the G/G model,''
  Nucl.\ Phys.\ B {\bf 408}, 345 (1993)
  [hep-th/9305010].

\bibitem{Smedback:2010ji}
  M.~Smedback,
  ``On thermodynamics of N=6 superconformal Chern-Simons theory,''
  JHEP {\bf 1004}, 002 (2010)
  [arXiv:1002.0841 [hep-th]].


\bibitem{Shenker:2011zf}
  S.~H.~Shenker and X.~Yin,
  ``Vector Models in the Singlet Sector at Finite Temperature,''
  arXiv:1109.3519 [hep-th].

\bibitem{Sachdev:1993pr}
  S.~Sachdev,
  ``Polylogarithm identities in a conformal field theory in three-dimensions,''
  Phys.\ Lett.\ B {\bf 309}, 285 (1993)
  [hep-th/9305131].

\bibitem{Petkou:1998wd}
  A.~C.~Petkou and M.~B.~Silva Neto,
  ``On the free energy of three-dimensional CFTs and polylogarithms,''
  Phys.\ Lett.\ B {\bf 456} (1999) 147
  [hep-th/9812166].

\bibitem{Christiansen:1999uv}
  H.~R.~Christiansen, A.~C.~Petkou, M.~B.~Silva Neto and N.~D.~Vlachos,
  ``On the thermodynamics of the (2+1)-dimensional Gross-Neveu model with complex chemical potential,''
  Phys.\ Rev.\ D {\bf 62} (2000) 025018
  [hep-th/9911177].

\bibitem{Gaiotto:2007qi}
  D.~Gaiotto and X.~Yin,
  ``Notes on superconformal Chern-Simons-Matter theories,''
  JHEP {\bf 0708}, 056 (2007)
  [arXiv:0704.3740 [hep-th]].

\bibitem{Feinberg:2005nx}
  J.~Feinberg, M.~Moshe, M.~Smolkin and J.~Zinn-Justin,
  ``Spontaneous breaking of scale invariance and supersymmetric models at finite temperature,''
  Int.\ J.\ Mod.\ Phys.\ A {\bf 20} (2005) 4475.

\bibitem{Yokoyama:2012fa}
  S.~Yokoyama,
  ``Chern-Simons-Fermion Vector Model with Chemical Potential,''
  arXiv:1210.4109 [hep-th].


\bibitem{Avdeev:1992jt}
  L.~V.~Avdeev, D.~I.~Kazakov and I.~N.~Kondrashuk,
  ``Renormalizations in supersymmetric and nonsupersymmetric nonAbelian Chern-Simons field theories with matter,''
  Nucl.\ Phys.\ B {\bf 391} (1993) 333.

\bibitem{Bardeen:1983rv}
  W.~A.~Bardeen, M.~Moshe and M.~Bander,
  ``Spontaneous Breaking of Scale Invariance and the Ultraviolet Fixed Point in O($n$) Symmetric $(phi^{6}$ in Three-Dimensions) Theory,''
  Phys.\ Rev.\ Lett.\  {\bf 52}, 1188 (1984).

\bibitem{Amit:1984ri}
  D.~J.~Amit and E.~Rabinovici,
  ``Breaking of scale invariance in phi**6 theory: tricriticality and critical end points,''
  Nucl.\ Phys.\ B {\bf 257}, 371 (1985).


\end{thebibliography}
\end{document}